\documentclass[12pt]{article}
\usepackage{amssymb} 
\usepackage{amsmath}
\usepackage{amsthm}   
\usepackage{bbm}
\usepackage{enumerate}

\usepackage{setspace}

\usepackage[english]{babel}
\usepackage{chapterbib}
\usepackage{xr}

\usepackage{bbold}	

\usepackage{graphicx}  
\usepackage{caption}[2008/08/24]
\usepackage{subcaption}

\usepackage{hyperref}

\usepackage{color}

\usepackage{natbib}

\theoremstyle{plain}
\newtheorem{Proposition}{Proposition}

\theoremstyle{remark}

\newcommand{\blind}{0}

\usepackage{algpseudocode}
\usepackage{algorithm}

\renewcommand{\Return}{\State \textbf{return} }
\algrenewcomment[1]{\(\hfill\backslash\backslash\) #1}

\newcommand{\R}{\mathbb{R}}

\newcommand{\ind}{\mathbbm{1}}

\newcommand\argmin{\operatornamewithlimits{argmin}}

\newcommand{\PCpluS}{\operatorname{PCpluS}}
\newcommand{\pcs}{\operatorname{PCpluS}}
\newcommand{\PELT}{\operatorname{PELT}}
\newcommand{\JIC}{\operatorname{JIC}}

\newcommand{\COPS}{\operatorname{COPS}}
\newcommand{\SOPS}{\operatorname{SOPS}}
\newcommand{\SCHACEcv}{\operatorname{SCHACE}}

\newcommand{\floor}[1]{\left\lfloor #1 \right\rfloor}

\begin{document}

\def\spacingset#1{\renewcommand{\baselinestretch}%
{#1}\small\normalsize} \spacingset{1}

\if0\blind
{
  \title{\bf Change-point regression with a\\ smooth additive disturbance}
  \author{Florian~Pein\\
    Lancaster University\\
    and\\
    Rajen~D.~Shah\thanks{
    The authors gratefully acknowledge support by the EPSRC grant EP/N031938/1 (StatScale). We thank Paul Fearnhead, Kate McDaid and Axel Munk for fruitful discussions.}\hspace{.2cm}\\
    University of Cambridge}
  \maketitle
} \fi

\if1\blind
{
  \bigskip
  \bigskip
  \bigskip
  \begin{center}
    {\LARGE\bf Change-point regression with a\\ smooth additive disturbance}
\end{center}
  \medskip
} \fi

\bigskip
\begin{abstract}
We assume a nonparametric regression model where the signal is given by the sum of a piecewise constant function and a smooth function. To detect the change-points and estimate the regression functions, we propose $\PCpluS$, a combination of the fused Lasso and kernel smoothing. In contrast to existing approaches, it explicitly uses the additive decomposition of the signal when detecting change-points. This is motivated by several applications and by theoretical results about partial linear model. We show how the use of the Epanechnikov kernel in the linear smoother results in very fast computation. Simulations demonstrate that our approach has a small mean squared error and detects change-points well. We also apply the methodology to genome sequencing data to detect copy number variations. Finally, we demonstrate its flexibility by proposing extensions to multivariate and filtered data. An R-package called \textit{PCpluS} is available on CRAN\footnote{\url{https://cran.r-project.org/package=PCpluS}}.
\end{abstract}

\noindent%
\vfill

\newpage
\spacingset{1.5} 

\section{Introduction}\label{sec:introduction}
Change-point regression is currently one of the most active research areas in Statistics. It has wide ranging applications, for instance in biochemistry \citep{hotz2013idealizing, pein2017heterogeneous}, climatology \citep{reeves2007review}, environmental analysis \citep{bolorinos2020consumption}, finance \citep{bai2003computation, kim2005structural}, genomics \citep{olshen2004circular}, medicine \citep{younes2014inferring}, network traffic data analysis \citep{lung2012distributed}, quality  monitoring \citep{d2011incipient} and speech processing \citep{harchaoui2009regularized}.

A simple but very common setting considered is that where observations $Y_1,\ldots,Y_n \in \R$ are modelled as
\begin{equation*}
Y_i = h^0(i/n) + \varepsilon_i,\quad i = 1,\ldots,n,
\end{equation*}
where $h^0$ is a piecewise constant signal, $n$ is the number of observations and $\varepsilon_1,\ldots,\varepsilon_n$ are random errors.

In many applications however, while the signal $h^0$ may change abruptly, between these change-points the signal may not be precisely constant. Examples include climatology \citep{QiuYandell02, Qiu03, WuZhao07}, finance \citep{Wang95, Gijbelsetal07, Lametal16, Liuetal18}, light transmittance \citep{abramovich2007estimation} and many more. We may model this via
\begin{equation}\label{eq:model}
	Y_i = f^0(i/n) + g^0(i/n) + \varepsilon_i,\quad i = 1,\ldots,n,
\end{equation}
where $f^0$ is piecewise constant with $f^0(0)=0$ (which we may assume without loss of generality), and $g^0$ is a smooth function representing a systematic disturbance. An example is given in Figure~\ref{fig:example}. An additive decomposition of this form is for instance plausible when the piecewise constant and the smooth component have different sources, as it is common in applications. For example, in genome sequencing data, copy number variations cause large abrupt changes while biases in the measurements are a source for systematic disturbances; see~Section \ref{sec:application}. A further example we study in Section~\ref{sec:filter} comes from patchclamp recordings: experiments that allow one to measure the conductance of a single transmembrane protein over time. Abrupt changes are of primary interest as they indicate for instance a blocking of the protein or a change of its conformation. However, base line fluctuations caused by small holes in the membrane or by vibrations can be seen as systematic disturbances.

Model \eqref{eq:model} may be contrasted with the more general model where the signal is only assumed to be piecewise smooth, i.e.\ all derivatives may change abruptly at a change-point location. Both models have been considered in the literature; however in cases where the additive decomposition in \eqref{eq:model} is assumed, it is only used in a post-processing step to re-estimate the smooth signal $g^0$ by smoothing the observations with the estimated piecewise constant signal subtracted, but not in the detection of the change-points themselves. Instead, for detecting change-points, the approaches developed involve applying established nonparametric smoothing procedures to segments between candidate changes. These include proposals based on kernel regression \citep{Mueller92, Qiu94, eubank1994nonparametric,  Qiu03, GijbelsGoderniaux04}, local polynomial regression \citep{Loader96, Spokoiny98, Gjbelsetal04, Gijbelsetal07, Mboupetal08, xia2015jump, Zhang16, wang2022data}, (smoothing) splines \citep{Koo97, Lee02, MiyataShen03, HuangQui10, YangSong14, Liuetal18, lee2021change, lu2023simultaneous}, wavelets \citep{Raimondo98, AntoniadisGijbels02, abramovich2007estimation} and nonparametric Bayesian approaches \citep{Denisonetal98, Ogden99, Dimatteoetal01, Punskayaetal02, Fearnhead05, Morenoetal13, Lee18}. A helpful review is given in \citet{qiu2005image}.

Whilst there may be benefits to using these methods when \eqref{eq:model} is heavily misspecified, their additional flexibility can sacrifice estimation accuracy of the change-points. To see this, consider a simplification of model \eqref{eq:model} where $f^0$ has a single change-point whose location is known. We may cast the problem as a partially linear model $Y_i = X_i \beta^0 + g^0(i / n) + \varepsilon_i$, where  $X := (0, \ldots, 0, 1, \ldots, 1)^\top  \in \R^{n \times 1}$ and $\beta^0 \in \R$ is the unknown jump size. Without using the additive structure of the model, our only option for estimating $\beta^0$ is to take the difference of the regression estimates to the right and left of the change-point analogously to the methods mentioned above. Our rate for estimating $\beta^0$ would then mirror the estimation error for nonparametrically estimating the regression functions, which would be strictly slower than the parametric $1/\sqrt{n}$ rate. On the other hand, classical methods for estimating $\beta^0$ tailored to the partially linear model setting can achieve
the optimal $1/\sqrt{n}$ rate. An example is the least squares estimator
\begin{equation}\label{eq:leastSquares}
\begin{split}
\hat{\beta} := & \argmin_{\beta \in \mathbb{R}} \| (I - S) (Y - X \beta)\|_2^2\\
=& \{X^\top (I - S)^\top  (I - S) X\}^{-1} X^\top  (I - S)^\top  (I - S) Y,\\
\text{ and } \hat{g}(i/n) :=& \{S(Y - X \hat{\beta})\}_i,
\end{split}
\end{equation}
of \citet{denby1984smooth}, see also \citep{speckman1988kernel}. Here, $ S \in \R^{n \times n}$ is a smoothing matrix (see \eqref{eq:NW} for a possible choice) and $Y := (Y_1,\ldots, Y_n)^\top $. The difference in rate aligns with the intuition that assuming the decomposition allows us to combine information about the smooth signal from the left and right sides of a potential change-point to improve estimation accuracy and detection power. In the context of change-point estimation, the only notable work in this direction is from  \citet{eubank1994nonparametric} who estimate the location and size of a single change-point in the derivative of an otherwise smooth function by using ideas from semiparametric statistics.

\subsection{Our contributions}
In this work, we consider the problem of estimating (potentially) multiple change-points in model \eqref{eq:model} and the smooth function $g^0$. Our approach is motivated by the least squares estimator of \eqref{eq:leastSquares}, and in particular makes use of a kernel smoothing matrix $ S$. However, it replaces the matrix $ X$ with one which encodes for all possible change-points, and includes a total variation regulariser for the associated coefficient vector similarly to the fused Lasso \citep{tibshirani2005sparsity}.
In this way, the estimator takes advantage of the additive structure in \eqref{eq:model} for better estimation accuracy.
As we explain in Section~\ref{sec:methodology}, the optimisation problem to be solved amounts to computing the Lasso where the coefficient vector has dimension $n - 1$. Naive computation can therefore be prohibitively costly for large $n$. The key bottleneck in the standard coordinate descent algorithm \citep{friedman2007pathwise} for computing the Lasso is the calculation of the dot product of each column of an $n \times (n-1)$ design matrix with specific columns indexed by a so-called active set. In general, for one such column in an active set, this would amount to an $\mathcal{O}(n^2)$ computation.
However as we show in Section~\ref{sec:kernelSmoothing}, when using the Epnechnikov kernel, this computation reduces to $\mathcal{O}(n \times h)$, where $h$ is the bandwidth of the kernel; often we will have $h \ll 1$.
This makes our method suitable for deployment on large-scale data.

To correct for the well-known effect of the total variation penalty tending to select too many change-points\footnote{In fact in our case, the inclusion of a smoothing matrix $ S$ in the least squares term as in \eqref{eq:leastSquares} acts as a form of preconditioner that mitigates this issue.} \citep{qian2016stepwise}, we propose further post-processing steps detailed in Section~\ref{sec:post-processing}. We refer to the final estimator as $\PCpluS$, standing for the \textbf{P}iecewise \textbf{C}onstant \textbf{plu}s Smooth Regression Estimator. Section~\ref{sec:simulation} demonstrates how $\PCpluS$ enjoys good empirical performance in a variety of settings and compares favourably in terms of average mean squared error and correct detection of change-points, to competing methods.
We also present an application to genome sequencing data to detect copy number variation in Section~\ref{sec:application}. In Section~\ref{sec:discussion} we discuss extensions of our method and in particular outline a variant for use with filtered data and an approach for multivariate settings. The supplementary material contains pseudocode, additional numerical results, and the proofs. 

\section{Methodology}\label{sec:methodology}
In the following, it will be convenient to treat our estimands $f^0$ and $g^0$ not as functions defined on $[0,1]$, but instead consider only their evaluations on $1/n, \ldots, n/n$. To this end, let $ f^* := (f^0(1/n), \ldots,f^0(n/n))^\top  \in \R^n$ and similarly for $g^* \in \R^n$.
To motivate our approach further, suppose first that the piecewise constant function $f^0$ in model \eqref{eq:model}, and hence $f^*$, is known.
Now $Y - f^*$ is a vector of observations in a vanilla nonparametric smoothing problem, to which we may apply standard procedures to estimate the smooth component $g^*$. For reasons that will become clear shortly, we specifically consider linear smoothers for this estimation problem, that is estimators of the form
\begin{equation}\label{eq:functionalG}
	\hat{g}(f)= S (Y - f),
\end{equation}
where $ S \in \R^{n \times n}$ is a smoothing matrix, $f \in \R^n$ is an arbitrary vector and $\hat{ g} \in \R^n$ estimates $g^*$. A popular example of a linear smoother is the Nadaraya--Watson estimator which in our case takes the form
\begin{equation}\label{eq:NW}
\hat{g}_j := \frac{\sum_{i=1}^n (Y_i - f_i) k\left(\frac{j-i}{nh}\right)}{\sum_{i=1}^n k\left(\frac{j-i}{nh} \right)}, \qquad \text{so } \;S_{ij} := \frac{k\left(\frac{j-i}{nh}\right)}{\sum_{l=1}^n k\left(\frac{l-i}{nh}\right)}.
\end{equation}
Here $k:\R \to [0,\infty)$ is a kernel and $h>0$ is a bandwidth.

%

On the other hand, were $g^*$ to be known, then $Y - g^*$ could be considered as observations in a standard piecewise constant change-point regression problem. Hence, $f^*$ may be estimated by any of the many existing change-point detection procedures, one of which is the fused Lasso \citep{tibshirani2005sparsity, lin2017sharp}:
\begin{equation}\label{eq:fusedLasso}
\hat{f} := \argmin_{f \in \R^n} \| (Y - g^*) - f\|_2^2 + \lambda \| f \|_{\operatorname{TV}}.
\end{equation}
Here $\lambda > 0$ is a tuning parameter,
and $\| f \|_{\operatorname{TV}} := \sum_{i=2}^n{\vert f_i - f_{i - 1} \vert}$ is the total variation norm.

In our situation both $f^*$ and $g^*$ are unknown, so we combine the two approaches above by replacing $g^*$ in \eqref{eq:fusedLasso} by the functional $\hat{g}(f)$ from \eqref{eq:functionalG}. Our estimator for $ f$ is then given by
\begin{equation}\label{eq:modifiedFusedLasso}
\begin{split}
\hat{f} := & \argmin_{f \in \R^n,\ f_1 = 0}  \Big\{ \| Y - \hat{g}(f) - f\|_2^2 + \lambda \| f \|_{\operatorname{TV}} \Big\} \\
 = & \argmin_{f \in \R^n,\ f_1 = 0} \Big\{ \| (I - S) (Y - f)\|_2^2 + \lambda \| f \|_{\operatorname{TV}} \Big\}.
\end{split}
\end{equation}
Here $I$ is the $n$-dimensional identity matrix and the constraint $f_1 = 0$ is added to ensure identifiability.
The use of a linear smoother and the fused Lasso, rather than taking arbitrary nonparametric smoothing and change-point detection procedures as starting points, results in a convex optimisation problem. In fact, \eqref{eq:modifiedFusedLasso} is simply a Lasso problem with response vector $\tilde{Y } := (I - S) Y$ and design matrix $\tilde{  X} := (I - S)  X$ where $ X \in \R^{n \times (n-1)}$ has $X_{ij} = \ind_{\{i>j\}}$. That is $\hat{f}_i= \hat{\beta}_i$ for $i \geq 2$ where $\hat{\beta}$ is given by
\begin{equation}\label{eq:Lasso}
	\hat{\beta} :=  \argmin_{ \beta \in \R^{n-1}} \left\{ \| \tilde{Y} - \tilde{X} \beta \|_2^2 + \|\beta\|_1 \right\}.
\end{equation}
We note that \citet[(2.7)]{norouzirad2022differenced} considers a Lasso problem of a similar form, but for an arbitrary design matrix and $(I - S)$ replaced by a differencing matrix.
While efficient solvers exist for Lasso problems, there are some computational challenges when simply applying these to solve the problem above. Firstly, when $n$ is large, as can certainly be the case in change-point detection problems, the design matrix will have $\mathcal{O}(n^2)$ entries. It may not be possible to store this in memory, and in any case computing all of the entries may already be an issue. Secondly, the optimisation \eqref{eq:modifiedFusedLasso} will typically need to be performed with a range of different smoothing matrices $ S$, each corresponding to different bandwidths or other tuning parameters, such that the final choice can be selected by cross-validation.
Furthermore, an issue with the estimator itself is that it may inherit from the fused Lasso, a tendency to over-estimate the number of change-points \citep{qian2016stepwise}. We turn first to the computational issues in Section~\ref{sec:kernelSmoothing} before discussing a post-processing procedure to tackle the potential over-selection of changes in Section~\ref{sec:post-processing}. Cross-validation, which particularly our setting is not entirely straightforward, is discussed in Section~\ref{sec:crossvalidation}.

%

\subsection{Computation} \label{sec:kernelSmoothing}
As discussed above, fast computation of the estimator $\hat{ f}$ \eqref{eq:modifiedFusedLasso} or equivalently $\hat{\beta}$ in \eqref{eq:Lasso} for an arbitrary smoothing matrix $ S$ presents serious computational challenges. It turns out however that when $ S$ is given by the Nadaraya--Watson estimator \eqref{eq:NW} with certain kernels, the computation can be dramatically sped up. Before we describe this, it will be helpful to briefly review how Lasso estimates are typically computed.

\subsubsection{Computation of Lasso solutions}
The fastest Lasso solvers use cyclic coordinate descent \citep{friedman2010regularization}, which involves repeatedly looping through the components of $\hat{\beta}$ and updating each in turn by optimising \eqref{eq:Lasso} over the single component keeping  all others fixed at their current values.
More precisely, we fix a decreasing sequence of $\lambda$-values $\lambda_0 > \lambda_1 > \cdots > \lambda_{n_{\lambda}}$ where $\lambda_0 := \max_{j} | \tilde{X}_{j}^\top \tilde{Y} |$; here and below, given a matrix $M$, $M_j$ denotes its $j$th column. By the Karush--Kuhn--Tucker (KKT) conditions, the Lasso solution $ \hat{\beta}^{(0)}$ at $\lambda_0$ will be $0\in \R^{n - 1}$.

We compute Lasso solutions at larger $\lambda$-values iteratively. Given a Lasso solution $\hat{\beta}^{(k)}$ at $\lambda_k$, we let the active set $\mathcal{A} \subseteq \{1,\ldots,n-1\}$ be the set of indices of  $\hat{\beta}^{(k)}$ that are non-zero. We first perform coordinate descent only among those components in the active set. This involves initialising $\hat{\beta}^{(k)} \gets\hat{\beta}^{(k-1)}$ and then performing the update
\begin{equation} \label{eq:soft_thresh}
\hat{\beta}^{(k)}_j \gets s_{\lambda_k} \left( \tilde{ X}_{j}^{\top}\tilde{Y} - \sum_{l \in \mathcal{A}, \, l \neq j} \tilde{X}_j^{\top} \tilde{X}_l \hat{\beta}^{(k)}_l \right),
\end{equation}
repeatedly cycling through $j \in \mathcal{A}$ until convergence. Here, $s_{\lambda}(x)$ is the soft-threshold operator $s_{\lambda}(x) := \operatorname{sign}(x)(\vert x \vert - \lambda)$. The KKT conditions certifying that $\hat{\beta}^{(k)}$ is a Lasso solution amount to checking
\begin{equation} \label{eq:KKT_check}
	\Big\lvert \tilde{ X}_{j}^{\top} \big(\tilde{Y}- \tilde{X} \hat{\beta}^{(k)}\big) \Big\rvert = \left| \tilde{ X}_{j}^{\top} \tilde{Y} -  \sum_{l \in \mathcal{A}} \tilde{X}_j^{\top} \tilde{X}_l \hat{\beta}^{(k)}_l \right| \leq \lambda_k
\end{equation}
for all $j \notin \mathcal{A}$.
Any components $j$ violating this inequality are added to the active set $\mathcal{A}$, and the coordinate descent steps \eqref{eq:soft_thresh} are performed once more\footnote{Further refinements to this basic procedure involving so-called `strong rules' can offer additional speed-ups \citep{tibshirani2012strong}; we also employ these in our implementation \citep{PCpluS}}. This is repeated until no violating components are present, at which point $\hat{\beta}^{(k)}$ will be a Lasso solution at $\lambda_k$ and we can proceed to the next value of $\lambda$. We see that the  key computational steps in \eqref{eq:soft_thresh} and \eqref{eq:KKT_check} involve calculating
\begin{equation*}
\tilde{X}_j^{\top} \tilde{Y} \qquad \text{and} \qquad \tilde{X}_j^{\top} \tilde{X}_l
\end{equation*}
for all $j$  and any $l$ appearing in an active set. We now describe how calculating these quantities can be made very fast when using the Epanechnikov kernel.

\subsubsection{Using the Epanechnikov kernel}
Recall that the Epanechnikov kernel is given by $k(x) := 3 (1 - x^2)_+ / 4$, where $(\cdot)_+$ denotes the positive part. One immediate benefit of using this kernel within the smoothing matrix $S$ comes from the fact that it is compactly supported. Let $L := \floor{nh}$ where $0<h \leq 1$ is the bandwidth, and define $k_l := k\big(l/(nh)\big)$ for $l=-L,\ldots,L$. Recall that we may write the smoothing matrix $S = DK$ where $K_{ij} =  k_{i-j} = k_{j-i}$ and $D \in \R^{n \times n}$ is diagonal with
\begin{equation} \label{eq:D_def}
D_{ii} := \left( \sum_{l=1}^n K_{il} \right)^{-1} = \left( \sum_{l=1 \vee (i-L) }^{n \wedge (i+L)} k_{l-i} \right)^{-1} = \left( \sum_{l=(1-i) \vee -L}^{(n-i) \wedge L} k_l \right)^{-1} .
\end{equation}
Thus $K$ is an $L$-banded Toeplitz matrix and $D$ has $\mathcal{O}(L)$ unique entries which may be computed in $\mathcal{O}(L)$ time. Furthermore, we have the following:
\begin{Proposition} \label{prop:X_tilde}
	The matrix $\tilde{X} = (I-S) X$ is $L$-banded and the $(n-2L) \times (n-1)$ submatrix of $\tilde{X}$ formed through excluding the first $L$ and last $L$ rows is Toeplitz. Moreover with $\mathcal{O}(L)$ pre-computation, any element of $\tilde{X}$ may be obtained in constant time.
\end{Proposition}

\begin{Proposition}\label{proposition:tildeXtildeX}
	We have that $\tilde{X}^\top \tilde{X}$ is $2L$-banded and the submatrix of $\tilde{X}^\top \tilde{X}$ formed through excluding the first $2L$ and last $2L$ rows is Toeplitz.
\end{Proposition}

Proofs are given Appendix~\ref{appendix:proofs}. Using these results alone, we find, for example, that computing $\tilde{Y}$ requires $\mathcal{O}(nL)$ operations as does computing $\tilde{X}_j^{\top} \tilde{Y}$ for all $j$ (with $\tilde{Y}$ computed, each of these dot products involves a sum of $\mathcal{O}(L)$ quantities). Similarly, computing the entries of $\tilde{X}^{\top} \tilde{X}$ would require $\mathcal{O}(L^3)$ operations since this involves $\mathcal{O}(L^2)$ dot products of pairs of vectors, each with $\mathcal{O}(L)$ non-zero entries. Importantly however, the parabolic form of the  Epanechnikov kernel allows one to speed these up by a further factor of $L$, so for example computing the dot products $\tilde{X}_j^{\top} \tilde{Y}$ for all $j$ requires only $\mathcal{O}(n)$ operations.
To see this, let $v \in \R^n$ be an arbitrary vector and first consider computing $K v$. For notational convenience, let us define $v_j := 0$ for all $j \notin \{1,\ldots,n\}$.

We have
\begin{align*}
	\frac{4}{3} (Kv)_i & =  \frac{4}{3} \sum_{l=1}^n K_{il} v_l = \sum_{l=i-L}^{i+L} \left\{1 - \left(\frac{l-i}{nh}\right)^2 \right\} v_l  \\
	&=  \sum_{l=i-L}^{i+L} v_l + \frac{1}{(nh)^2} \sum_{l=i-L}^{i+L} (l-i)^2 v_l =: a_i + c_i.
\end{align*}
Next also defining $b_i := \sum_{l= i-L}^{i+L} (i-l)v_l$, we see that we have the recursions
\begin{align*}
	a_{i + 1} & = a_i + v_{i + L + 1} - v_{i - L},\\
	b_{i + 1} & = b_i + c_i + v_{i + L + 1} (L + 1) - v_{i - L} L,\\
	c_{i + 1} & = c_i + 2 b_i + a_i + v_{i + L + 1} (L + 1)^2 - v_{i - L} L^2.
\end{align*}
Hence we can compute all $(a_1,b_1,c_1),\ldots,(a_n,b_n,c_n)$ in $\mathcal{O}(n)$ operations, and thus the computation of $Kv$ is $\mathcal{O}(n)$. Thus computing $\tilde{Y} = (I-DK)Y$ is an $\mathcal{O}(n)$ operation. Moreover,
\[
\tilde{X}^{\top} v = X^{\top} (I - S^{\top}) v = X^{\top} (I - DK) v.
\]
So, if $v$ has only a contiguous block of $\mathcal{O}(L)$ non-zero entries, then computing  $\tilde{X}^{\top} v$ will only have $\mathcal{O}(L)$ cost.
Thus computing an entire column of $\tilde{X}^{\top} \tilde{X}$ is only an $\mathcal{O}(L)$ operation. Runtime efficiency of this algorithm is demonstrated by numerical experiments in Section~\ref{sec:runtime}.

\subsection{Post-processing}\label{sec:post-processing}
\begin{figure}[htb]
  \begin{subfigure}[t]{0.49\linewidth}
	    \centering
	    \includegraphics[width=\linewidth]{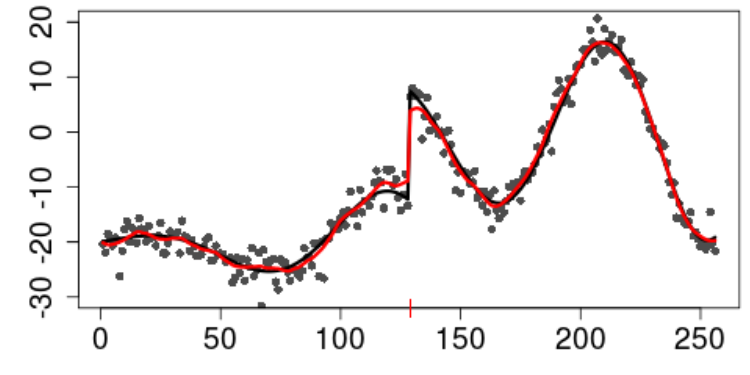} 
	    \caption{without post-processing}
	    \label{fig:exampleFirstStep} 
	  \end{subfigure}
  \begin{subfigure}[t]{0.49\linewidth}
	    \centering
	    \includegraphics[width=\linewidth]{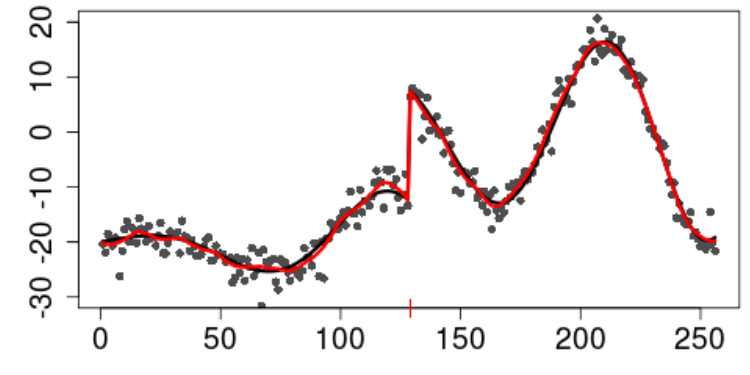} 
	    \caption{after post-processing}
	    \label{fig:exampleFit} 
	  \end{subfigure}
  \caption{An illustrative example of a signal composed of a sum of piecewise constant and smooth functions (black line), and simulated observations around this (grey points). $\PCpluS$ (red line) even without post-processing (left panel) correctly identifies the change-point and estimates the signal with high accuracy though the size of the change is underestimated; this is corrected in our final $\PCpluS$ algorithm (right panel).}
  \label{fig:example} 
\end{figure}

While the fit provided by the estimator proposed above works fairly well (see Figure~\ref{fig:exampleFirstStep}), in general we observe  two systematic flaws. Firstly, sizes of change-points tend to be underestimated because of the shrinkage effect of the total variation penalty in \eqref{eq:modifiedFusedLasso}. Secondly, false positives of small jump size are detected, particularly when the bandwidth is large. This latter deficiency, inherited from the fused Lasso, is slightly less severe than might be expected since when the bandwidth is small, the matrix $ S$ acts as a form of preconditioner in the fused Lasso problem; such preconditioning has been shown to mitigate the issue of false positives with the fused Lasso in \citet{qian2016stepwise}. However, particularly for larger bandwidths, this problem remains. To deal with these issues, we propose two post-processing steps. 

%
%
%
%

To correct for false positives we re-estimate the change-points in a first post-processing step by applying $\PELT$ \citep{killick2012optimal} to the observations minus the estimated smooth signal, that is $Y - \hat{g}$. This is motivated by the fact, that provided the smooth signal is estimated reasonably well,  $Y - \hat{g} \approx Y - g^*$ 
can be treated as a piecewise constant regression problem. More precisely, we re-estimate $f$ by
\begin{equation}\label{eq:pelt}
\tilde{f} := \argmin_{f \in \R^n} \| (Y - \hat{g}) - f\|_2^2 + \lambda_{\PELT} \| f \|_0,
\end{equation}
where $\| \cdot \|_0$ denotes the number of change-points and $\lambda_{\PELT}:=2\sigma_0\log(n)$ denotes the $\operatorname{SIC}$-penalty. Here $\sigma_0^2$ is the variance of the errors when it is known; otherwise this is replaced by its estimate $\operatorname{IQR}\left(Y_{2}-Y_1,\ldots, Y_n-Y_{n-1}\right)/\{2\sqrt{2}\Phi^{-1}(0.75)\}$, where
$\Phi^{-1}$ denotes the quantile function of the standard Gaussian distribution.



As discussed above, a second issue with the fused Lasso penalty is that it tends to result in underestimation of the jump sizes. To handle this, in a 
second post-processing step we repeat \eqref{eq:modifiedFusedLasso} without penalisation but with changes restricted to the change-point locations estimated in $\tilde{f}$. This step is equivalent to \eqref{eq:leastSquares}, but with (potentially) multiple change-points encoded by multiple regressors. More precisely, let $0<\hat{\tau}_1<\cdots<\hat{\tau}_{\hat{K}}<1$ be the change-points estimated in $\tilde{f}$ and set $\hat{J}:=\{n\hat{\tau}_1,\ldots,n\hat{\tau}_{\hat{K}}\}$. Our final `$\PCpluS$' estimators for $f$ and $g$ are given by
\begin{equation}\label{eq:LSpost-processing}
\hat{f}^{\mathrm{PC}+} := \argmin_{\substack{f \in \R^n,\ f_1 = 0,\\ f_i\neq f_{i+1}\ \Leftrightarrow\ i \in \hat{J}}} \|(I - S) (Y - f)\|_2^2,
\end{equation}
with $\hat{g}^{\mathrm{PC}+} := S (Y - \hat{f}^{\mathrm{PC}+})$. 
The optimisation problem \eqref{eq:LSpost-processing} is an ordinary least squares regression problem with regression matrix $\tilde{X}_{\hat{J}}$, where $\tilde{X}_{\hat{J}} \in \R^{n \times |\hat{J}|}$ denotes the sub-matrix of $X$ consisting of those columns indexed by $\hat{J}$. Since $|\hat{J}|$ is often small, and also since from Proposition~\ref{proposition:tildeXtildeX}, $(X_{\hat{J}} ^{\top} X_{\hat{J}})_{ij} = 0$ if $n\vert \hat{\tau}_i - \hat{\tau}_j\vert > 2L$, this step is very fast.
The full $\PCpluS$  procedure is summarised in Algorithm~\ref{alg:pcplus} in Appendix~\ref{appendix:pseudocode}.

An illustrative $\PCpluS$ fit is shown in Figure~\ref{fig:exampleFit}. We see that $\PCpluS$ is able to detect the change-points 
correctly and estimates the size of the change and the smooth function well.

\subsection{Cross-validation}\label{sec:crossvalidation}
There are two tuning parameters to be chosen for $\PCpluS$: the bandwidth of the smoothing matrix $ S$ and the fused Lasso penalty $\lambda$ (note that we use the same bandwidth in \eqref{eq:modifiedFusedLasso} and \eqref{eq:LSpost-processing} to simplify the selection and to reduce computation time).
We propose to select these via $2$-fold cross-validation with time ordered folds, see \citep[(4)]{pein2021crossvalidation}.
We also advocate the use of $L_1$-loss instead of the standard $L_2$-loss in the cross-validation criterion. This follows the recommendation from \citet{pein2021crossvalidation}, where it is shown that using an $L_2$-loss criterion with cross-validation in simple piecewise constant change-point regression can result in systematic under- or over-selection of the number of change-points, and poor predictive performance in settings where changes are easily detectable. In fact these issues are even more severe in our setting, as we now explain.

Consider a piecewise constant signal with a single large change-point occurring near the middle of the data with large jump size $\Delta$. Clearly, a large bandwidth, ideally infinitely large, would be desirable here since there is no smooth function component to the signal. However, in this simple setting, cross-validation with $L_2$-loss is likely to select a very small bandwidth. If the point following the large change is in the hold-out set, this will be wrongly estimated to have the mean of the preceding observations and incur an error of roughly $\Delta^2$. On the other hand, when a very small bandwidth is used, the out-of-sample prediction will be the average of the observations either side of the change-point, which will then incur an error of roughly $\Delta^2/2$. Thus in total, we can expect that the cross-validation criterion will be at least $\Delta^2$ for the piecewise constant regression, and $\Delta^2/2$ plus an additional contribution from the higher variance of the regression function estimate due to the small bandwidth. The result is that in what should be a straightforward setting where $\Delta^2$ is very large, the former can dominate the latter, and result in cross-validation incorrectly selecting a small bandwidth.

The cross-validation criterion is optimised by a grid search. For the bandwidth we use an exponential grid from $2.01 / n$ to $0.5$ with by default $n_h:=30$ entries plus the value infinity, for which the whole estimator simplifies to $\PELT$ \citep{killick2012optimal}. For each bandwidth, an exponential grid with by $n_\lambda := 30 $ entries is searched for the penalty $\lambda$. Out-of-sample predictions for cross-validation can be computed in the same time complexity using the principles outlined the previous sections, though detailed formulas differ slightly. 

\section{Simulations}\label{sec:simulation}
In this section we investigate the performance of $\PCpluS$ in three  simulations. In Section \ref{sec:simulationSmoothPlusCp} we consider the setting from \citet{abramovich2007estimation} which consists of four signals that can be decomposed into a piecewise constant plus a smooth function. In Section \ref{sec:simulationSmoothartefacts} we look at the setting from \citet{olshen2004circular} which considers a piecewise constant function plus smooth artefacts. Finally, we investigate in Section \ref{sec:simulationRobustness} the robustness of $\PCpluS$ to violations of the assumption in \eqref{eq:model} that the signal can be decomposed into a piecewise constant plus a smooth function. All simulations are performed in \texttt{R} and repeated $10\,000$ times. Code for the simulations and intermediate results are available on Github\footnote{\url{https://github.com/FlorianPein/simulations_PCpluS}}.

\subsubsection*{Methods}
For $\PCpluS$ we call the function \textit{pcplus} with the tuning parameters, the fused Lasso penalty $\lambda$ and the bandwidth $h$, selected by cross-validation, for which we call the function \textit{cv.pcplus}. Both functions are available in the \textit{R}-package \textit{PCpluS}. 

We compare our approach with the cross-validation procedures $\COPS$ and $\SOPS$ from \citep{wang2022data} for which we call the \textit{COPS} and \textit{SOPS} functions in the \textit{R}-package \textit{jra} with their default parameters. Both provide only an estimation for the change-points, but not for the regression function, so we cannot report mean squared errors. Secondly, we include $\SCHACEcv$ from the PhD thesis \citep{lu2023simultaneous} for which we call the function \textit{main.SCHACE} in the \textit{R}-package \textit{SCHACE} with the cross-validation option. This approach cites an earlier version of this paper and follows a suggestion therein that kernel smoothing in our current approach may be replaced by smoothing splines. Unfortunately, the implementation is very slow, and takes several seconds for $n = 256$ data points; hence we only include it in the simulations in Section~\ref{sec:simulationSmoothPlusCp} where $n = 256$ and repeated the simulations only 100 times. Furthermore, we used our own implementation of the method from \citet{xia2015jump}, which is based on a jump information criterion and hence abbreviated by $\JIC$ in the following. We follow the authors' recommendation to use a  bandwidth of $0.3 n^{-1/5}$. A systematic comparison with further existing piecewise smooth regression approaches is difficult, since to the best of our knowledge software is not available for most existing approaches.
This includes $\operatorname{ABS1}$ from \citet{abramovich2007estimation}, an approach that selects tuning parameters by generalised cross-validation. We nevertheless report their results where available ($n = 256$ and $a$ equal to $6$ or $4$ in Section~\ref{sec:simulationSmoothPlusCp}).
Finally, we have included the piecewise constant regression approach $\PELT$ \citep{killick2012optimal} for which we call the \textit{cpt.mean} function in the \textit{change-point} package with the \textit{"SIC"} penalty. The standard deviation is once again estimated in advance by an IQR-type estimator based on first order differences, see Section~\ref{sec:post-processing}.


\subsubsection*{Evaluation}
As well as evaluating the estimation of the signal via the average mean squared error (MSE), we also assess the estimation of the number of true change-points $K$. For an estimate $\hat{K}$, we
report the bias $\hat{K} - K$, how often $\hat{K} - K$ is $<-2$, equal to $-2,\ -1,\ 0,\ 1,\ 2$, and $>2$, as well as the average percentage that a true change-point is detected (averaged about all true changes). To this end, we say a change-point is detected if the distance between the true change-point location and a detected change-point is less than a given tolerance. For this analysis we have chosen the tolerance to be the minimum between $3$ and the smallest distance between two change-points divided by two (which is $2$ for the block signal). Note that for reasons of clarity and comprehensibility we sometimes omit some methods when evaluating change-point detection accuracy if they are perform too poorly or any not designed for change-point detection.

\subsection{Smooth plus piecewise constant signal}\label{sec:simulationSmoothPlusCp}
\begin{figure}[htb]
  \begin{subfigure}[t]{0.49\linewidth}
    \centering
    \includegraphics[width=\linewidth]{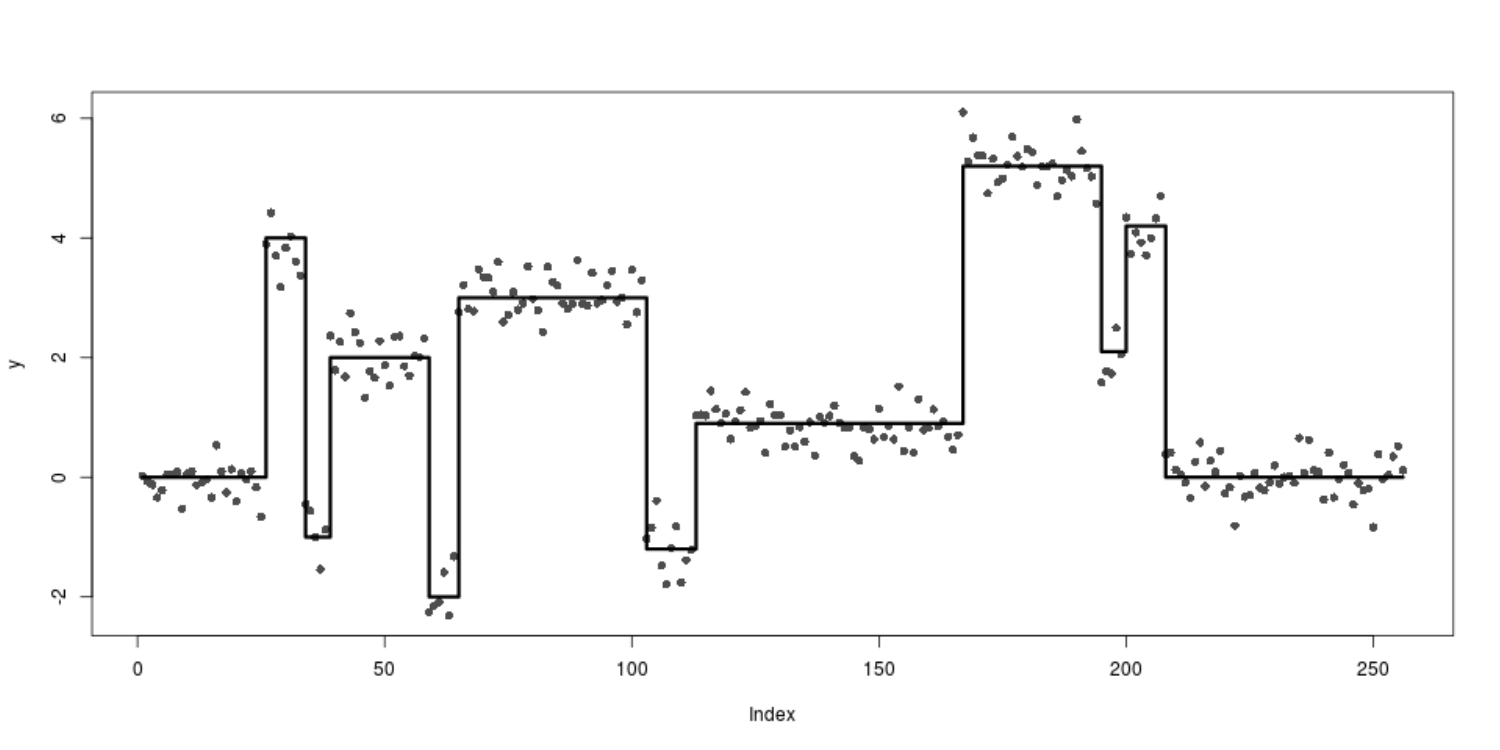} 
    \caption{blocks}
    \label{fig:blocks} 
  \end{subfigure}
  \begin{subfigure}[t]{0.49\linewidth}
    \centering
    \includegraphics[width=\linewidth]{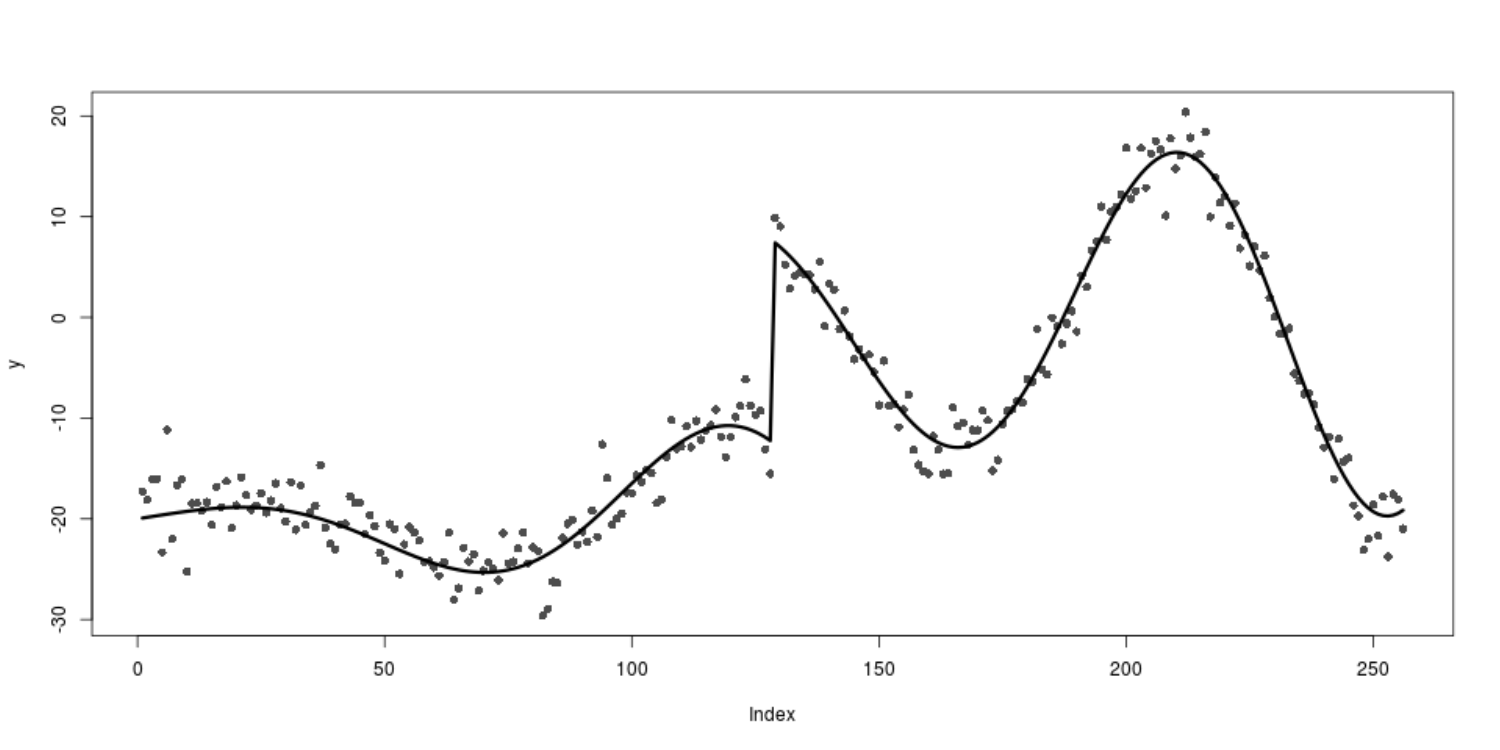} 
    \caption{burt}
    \label{fig:burt} 
  \end{subfigure}
  \begin{subfigure}[t]{0.49\linewidth}
    \centering
    \includegraphics[width=\linewidth]{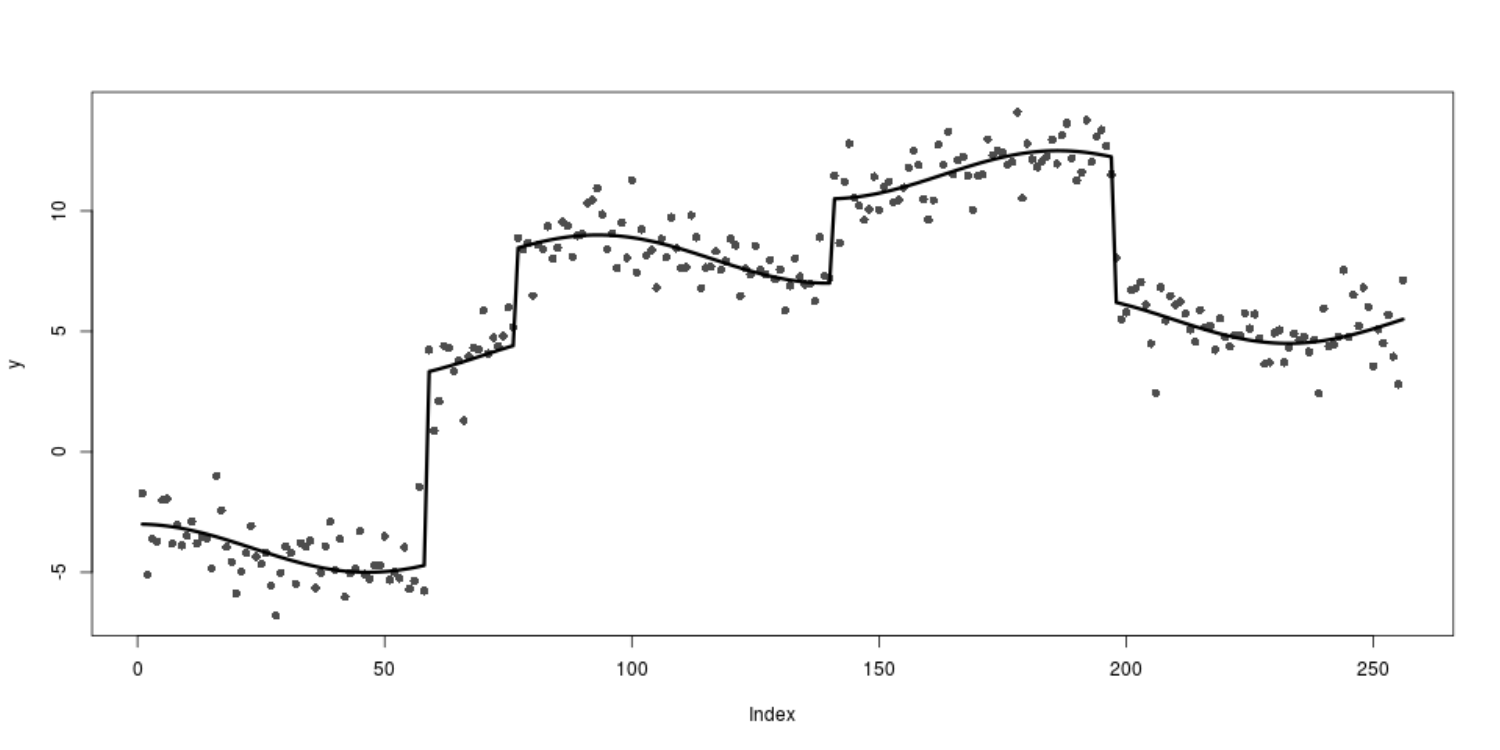} 
    \caption{cosine}
    \label{fig:cosine} 
  \end{subfigure}
  \begin{subfigure}[t]{0.49\linewidth}
    \centering
    \includegraphics[width=\linewidth]{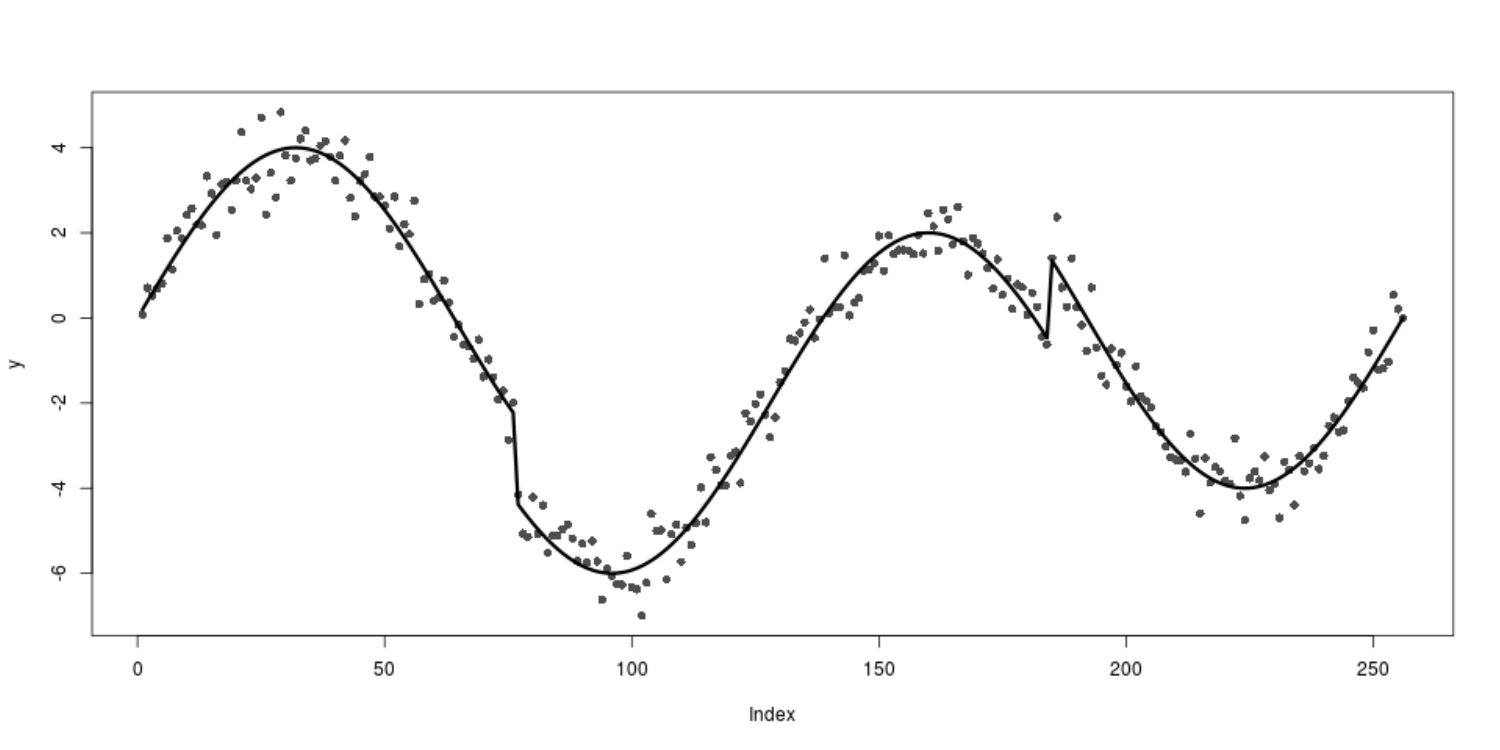} 
    \caption{heavisine}
    \label{fig:heavisine} 
  \end{subfigure}
  \caption{Visualisation of the piecewise constant plus smooth signals.}
  \label{fig:4signals} 
\end{figure}

In this section we repeat and extend the simulations study from \citet{abramovich2007estimation} who consider four different signals: blocks, burt, cosine, heavisine. These signals are visualised in Figure \ref{fig:4signals} and a formal definition can be found in their publication. All of these signals satisfy our model \eqref{eq:model} of a piecewise constant function plus a smooth function. Most of them involve a significant smooth component and hence we will focus mainly on the averaged mean squared error to assess the estimations, but also look briefly at the estimation of the change-points.

The standard deviation of the error is set to the value obtained by applying the empirical standard deviation formula to the signal and dividing by $a\in \{1,2,4,6\}$. \citet{abramovich2007estimation} considered $n=256$ and $a=6$ and $a=4$. In our simulations we are extending this to $n = 256$ with $a$ equal to $6,\ 4$ and $2$ as well as $n = 1024$ with $a$ equal to $4,\ 2$ and $1$. 

Results for $n=256$ and $a=4$ about the MSE are shown in Table \ref{tab:aMSEn256a4}. Results for other $n$ and $a$ are shown in Tables \ref{tab:aMSEn256a6}--\ref{tab:aMSEn1024a1} in Appendix \ref{sec:appendixSimulationResults} to keep this section readable. Results about the detection of change-points for $n=256$ and $a=4$ are shown in Table \ref{tab:CPn256a4} in Appendix \ref{sec:appendixSimulationResults}. We omit the results for other $n$ and $a$, since they allow qualitatively the same comparison between the methods and all methods improve rather similarly with increasing $n$ and $a$ (decreasing noise). 

\begin{table}[ht]
\centering
\begin{tabular}{lcccc}
  \hline
Method & blocks & burt & cosine & heavisine \\ 
  \hline
$\pcs$ & \textbf{0.02423} & 1.062 & \textbf{0.2819} & 0.09191 \\ 
  $\JIC$ & 1.833 & 23.84 & 0.5638 & 1.588 \\ 
   $\operatorname{ABS1}$ & 1.52 & 0.87 & 0.33 & 0.35\\ 
  $\PELT$ & 0.05983 &  5.87 & 0.7236 & 0.3701 \\ 
    $\SCHACEcv$ & 0.04219 & \textbf{0.6355} & 0.3409 & \textbf{0.07003} \\ 
   \hline
\end{tabular}
\caption{MSEs for $n = 256$ and $a = 4$.} 
\label{tab:aMSEn256a4}
\end{table}

We find in Tables~\ref{tab:aMSEn256a4}~and~\ref{tab:aMSEn256a6}-\ref{tab:aMSEn1024a1} that $\PCpluS$ is the method with the smallest averaged mean squared error or close to the best method. It outperforms other piecewise smooth regression approaches (JIC and ABS1) with the exception of $\operatorname{ABS1}$ for the burt signal and $\SCHACEcv$ for burt and heavisine. Remarkably, $\PCpluS$ even achieves a smaller MSE than PELT for the blocks signal, a piecewise constant regression problem for which $\PELT$ is designed, particularly when the noise level is small. This can be explained by the fact that we use cross-validation to select our parameters while the $\PELT$ penalty aims primarily for a correct estimation of the change-points. 

Table~\ref{tab:CPn256a4} shows that $\PCpluS$ detects change-points well, only the change-points in the heavisine signal are often missed. However, it also shows a tendency to slightly overestimate the number of change-points. Nonetheless, $\JIC$ outperforms our method only for the cosine signal, a signal which favours its bandwidth choice. Overall, the simulation clearly shows that a default bandwidth does not lead to good results and tuning parameter selection is still a challenging research topic when one aims to detect change-points in a piecewise smooth signal well. Unsurprisingly, $\PELT$ outperforms $\PCpluS$ when the signal is piecewise constant but heavily overestimates the number of change-point when it is not. $\COPS$ performs well unless the data is piecewise constant. It outperforms our estimator for the burt and the heavisine signal. $\SOPS$ systematically underestimates the number of change-points. Finally, $\SCHACEcv$ suffers from overestimations, but works well for the burt signal. 

Overall, the comparisons do not vary much with the number of observations $n$ and in the signal-to-noise ratio $a$. Furthermore, it depends on the signal which method performs well. It seems that $\PCpluS$ works well when the gradient does not change too substantially, but is outperformed by other methods for signals with large gradient changes such as burt and heavisine, though still showing a decent performance.

\subsection{Run time}\label{sec:runtime}
Here we examine the run time of $\PCpluS$. The algorithm is implemented in \texttt{R} and a software package is available on CRAN \citep{PCpluS}. The run time intensive code is written in C++ and interfaced. The overall computation of $\PCpluS$ is dominated by the computation of \eqref{eq:modifiedFusedLasso}; hence we focus on this aspect in our numerical experiments. We compare our implementation to the highly successful \textit{glmnet} \citep{friedman2010regularization} for solving the Lasso. We supply \textit{glmnet} with the matrix $\tilde{X}$ which we compute using C++ code that makes use of the banded and nearly Toeplitz structure of the matrix (Proposition~\ref{prop:X_tilde}); a version that computed this naively would be significantly slower. Note that \textit{glmnet} is able to exploit the sparsity in predictor matrices to speed-up computation.

We compare the two methods on one of the simulations from the previous section involving the cosine signal with the standard deviation of the noise governed by the parameter $a=4$ with varying numbers of observations $n$. We compare the runtime of the cross-validation and of the estimation procedure, where for the latter tuning parameters are obtained by running cross-validation once. All results are averaged over 100 repetitions. We see that our implementation is substantially faster than that given by  \textit{glmnet}, and the magnitude of the improvement increases with $n$.

\begin{figure}[htb]
  \begin{subfigure}[t]{0.49\linewidth}
    \centering
    \includegraphics[width=\linewidth]{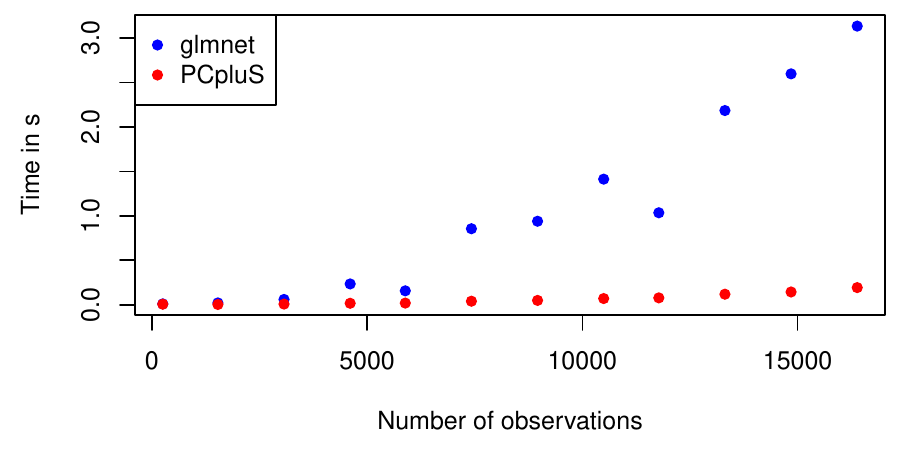} 
    \caption{Estimation}
    \label{fig:timeEst} 
  \end{subfigure}
  \begin{subfigure}[t]{0.49\linewidth}
    \centering
    \includegraphics[width=\linewidth]{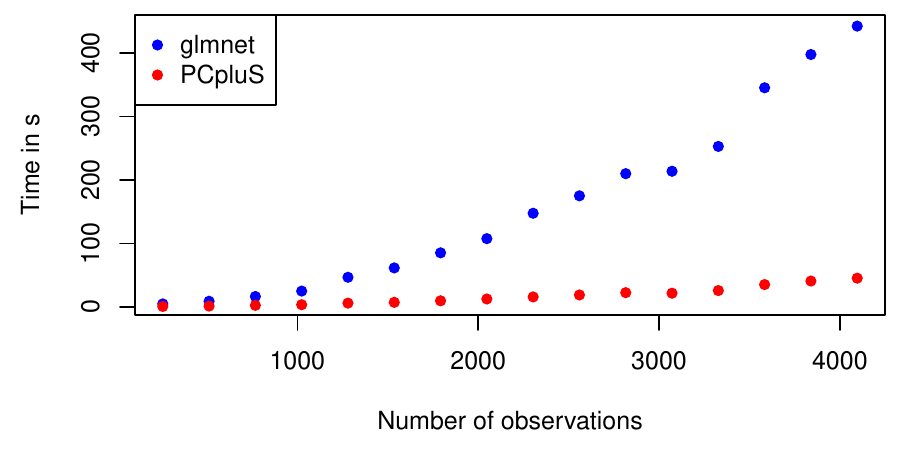} 
    \caption{Cross-validation}
    \label{fig:timeCV} 
  \end{subfigure}
  \caption{Runtime comparison between our implementation \textit{PCpluS} and an implementation using \textit{glmnet}.}
  \label{fig:time} 
\end{figure}

\subsection{Piecewise constant signal plus smooth artefacts}\label{sec:simulationSmoothartefacts}
In this section we once more consider signals that can be decomposed into a piecewise constant function and a smooth function. However, the focus here is on the piecewise constant component while the smooth component has a rather small amplitude, i.e.\ one might see it as a (small) `deterministic noise' contribution to an
otherwise piecewise constant signal.

\begin{figure}[!htb]
\centering
\includegraphics[width = 0.8\textwidth]{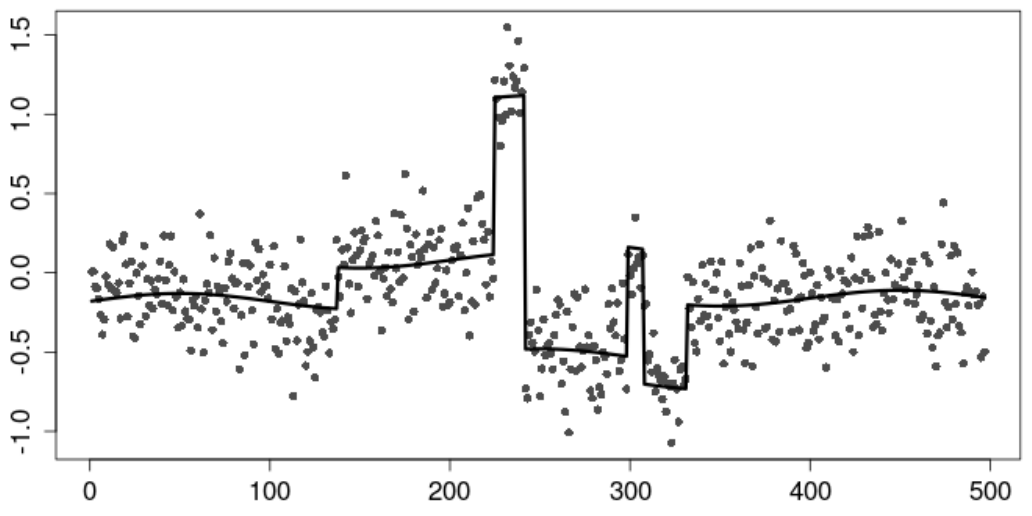}
\caption{The piecwise constant signal plus smooth artefacts with $a=0.01$ and $b = 0.2$ (black line) and exemplary observations (grey points).}
\label{fig:signalCpLong}
\end{figure}

To this end, we revisit and extend the simulation study from \citet{olshen2004circular}. Motivated by genome sequencing, these authors investigated the robustness of change-point approaches to smooth artefacts. This setting was later picked up by several piecewise constant change-point papers, e.g.\ \citet{zhang2007modified, niu2012screening, frick2014multiscale}. We simulate $n=497$ observations with standard deviation $0.2$. The piecewise constant signal has six change-points at locations $(138, 225, 242, 299, 308, 332) / n$ and function levels $(-0.18, 0.08, 1.07, -0.53, 0.16, -0.69, -0.16)$. The smooth artefacts are given by $0.25 b \sin(a \pi x n)$. We vary $b\in \{0, 0.2, 0.8\}$ and $a\in \{0.01, 0.025\}$ (long and short artefacts / trends). Note that $b=0.8$ was not considered in previous studies. An example for $a=0.01$ and $b=0.2$ is presented in Figure \ref{fig:signalCpLong}. Simulation results are given in Tables~\ref{tab:aMSEcp} and~\ref{tab:CPcp}.\\

\begin{table}[ht]
\centering
\begin{tabular}{lccccccc}
  \hline
 & $a = 0,$ & 
 $a = 0.01,$ & $a = 0.025,$ & $a = 0.01,$ & $a = 0.025,$ \\
Method & $b = 0$ & $b = 0.2$ & $b = 0.2$ & 
 $b = 0.8$ & $b = 0.8$ \\ 
  \hline
$\pcs$ & 0.001801 & \textbf{0.002287} & \textbf{0.002868} & 
 \textbf{0.003031} & \textbf{0.004358} \\ 
  $\JIC$ & 0.03389 & 0.03346 & 0.03637 & 
   0.03671 & 0.05645 \\ 
 $\PELT$ & \textbf{0.001564} & 0.002556 & 0.002884 & 
  0.00589 & 0.009259 \\ 
   \hline
\end{tabular}
\caption{Averaged mean squared errors for the piecewise constant signal plus smooth artefacts for various $a$ and $b$.} 
\label{tab:aMSEcp}
\end{table}

\begin{table}[ht]
\centering
\begin{tabular}{cclccccccccc}
  \hline
a & b & Method & $\hat{K} - K$ & $<-2$ & $-2$ & $-1$ & $0$ & $1$ & $2$ & $> 2$ & \% detected \\ 
  \hline
    0 &     0 & $\pcs$ & 0.2132 &  0.14 &  0.67 &  8.51 & 68.14 & 16.01 &  4.95 &  1.58 & 93.05 \\ 
      0 &     0 & $\JIC$ & -4.272 & 98.41 &  1.58 &  0.01 &     0 &     0 &     0 &     0 & 28.45 \\ 
      0 &     0 & $\PELT$ & 0.0766 &     0 &     0 &     0 & \textbf{93.43} &   5.6 &  0.87 &   0.1 &  97.3 \\ 
      0 &     0 & $\COPS$ & -3.575 & 89.29 &  6.45 &  2.55 &  1.05 &   0.4 &  0.14 &  0.12 & 34.86 \\ 
      0 &     0 & $\SOPS$ & -4.042 &  68.8 &     0 & 14.24 &  7.48 &  4.26 &   1.9 &  3.32 & 13.36 \\ 
   \hline
 0.01 &   0.2 & $\pcs$ & 0.2427 &  0.19 &  0.87 & 17.39 & 49.53 & 23.62 &  6.21 &  2.19 & 90.27 \\ 
   0.01 &   0.2 & $\JIC$ & -4.115 & 98.15 &  1.79 &  0.06 &     0 &     0 &     0 &     0 &    31 \\ 
   0.01 &   0.2 & $\PELT$ & 0.4548 &     0 &     0 &     0 & \textbf{63.08} & 29.67 &  6.16 &  1.09 & 95.73 \\ 
   0.01 &   0.2 & $\COPS$ & -3.571 & 89.59 &  6.42 &  2.52 &  0.88 &  0.37 &  0.14 &  0.08 & 35.17 \\ 
   0.01 &   0.2 & $\SOPS$ & -3.945 & 66.99 &     0 & 15.15 &  8.36 &  4.32 &   1.9 &  3.28 & 14.06 \\ 
   \hline
0.025 &   0.2 & $\pcs$ & 0.508 &  0.29 &  1.58 & 11.27 & 46.26 & 24.12 & 11.23 &  5.25 & 89.19 \\ 
  0.025 &   0.2 & $\JIC$ & -4.418 & 98.49 &  1.47 &  0.04 &     0 &     0 &     0 &     0 & 26.05 \\ 
  0.025 &   0.2 & $\PELT$ & 0.2841 &     0 &     0 &     0 & \textbf{79.19} & 15.05 &  4.47 &  1.29 & 95.58 \\ 
  0.025 &   0.2 & $\COPS$ & -3.474 & 85.75 &  8.84 &  3.11 &  1.39 &  0.55 &  0.21 &  0.15 & 34.95 \\ 
  0.025 &   0.2 & $\SOPS$ & -3.76 & 64.58 &     0 & 15.22 &  8.89 &  5.24 &  2.32 &  3.75 & 15.08 \\ 
   \hline
 0.01 &   0.8 & $\pcs$ & 0.3985 &  0.43 &   2.3 & 23.57 & \textbf{34.21} & 22.66 &  9.56 &  7.27 & 84.28 \\ 
   0.01 &   0.8 & $\JIC$ & -4.11 & 98.16 &  1.84 &     0 &     0 &     0 &     0 &     0 & 31.03 \\ 
   0.01 &   0.8 & $\PELT$ & 3.614 &     0 &     0 &     0 &     0 &  0.63 & 12.09 & 87.28 & 92.03 \\ 
   0.01 &   0.8 & $\COPS$ & -3.572 & 89.89 &     6 &  2.59 &  0.88 &  0.34 &  0.17 &  0.13 & 35.62 \\ 
   0.01 &   0.8 & $\SOPS$ & -3.823 & 65.12 &     0 & 16.31 &  8.48 &  4.51 &  1.84 &  3.74 & 14.87 \\ 
   \hline
0.025 &   0.8 & $\pcs$ & 0.1707 &  3.07 & 12.94 & 31.72 & \textbf{20.13} & 14.53 &  7.33 & 10.28 & 76.28 \\ 
  0.025 &   0.8 & $\JIC$ & -4.891 & 99.65 &  0.35 &     0 &     0 &     0 &     0 &     0 & 18.43 \\ 
  0.025 &   0.8 & $\PELT$ & 9.299 &     0 &     0 &     0 &     0 &     0 &  0.01 & 99.99 & 88.25 \\ 
  0.025 &   0.8 & $\COPS$ & -3.102 & 78.19 &  9.91 &  5.17 &  2.84 &  1.82 &  1.32 &  0.75 & 36.34 \\ 
  0.025 &   0.8 & $\SOPS$ & -1.995 & 42.69 &     0 & 16.86 &  13.2 &   9.4 &   6.2 & 11.65 & 24.53 \\ 
   \hline
\end{tabular}
\caption{Summary results about the detection of change-points for the piecewise constant signal plus smooth artefacts for various $a$ and $b$.} 
\label{tab:CPcp}
\end{table}

Tables~\ref{tab:aMSEcp}~and~\ref{tab:CPcp} reveal that $\PCpluS$  performs very well in this setting with respect to the 
MSE
 and also reasonably well in terms of detecting change-points correctly. Moreover, its advantage over the other methods tends to increase as the size of the artefacts are increased. Without artefacts $\PELT$ has a slightly smaller averaged mean squared error, but in the presence of artefacts $\PCpluS$ is equal when the artefacts are small or preferable when the artefacts are larger. Similarly, when looking at the detection of change-points $\PELT$ performance a bit better without artefacts or when the artefacts are small, but is worse when the artefacts are larger. Hence, as expected $\PCpluS$ is much less affected by such artefacts, but they still worsens results. All other competitors are clearly outperformed.

\subsection{Robustness}\label{sec:simulationRobustness}
\begin{figure}[!htb]
\centering
\includegraphics[width = 0.8\textwidth]{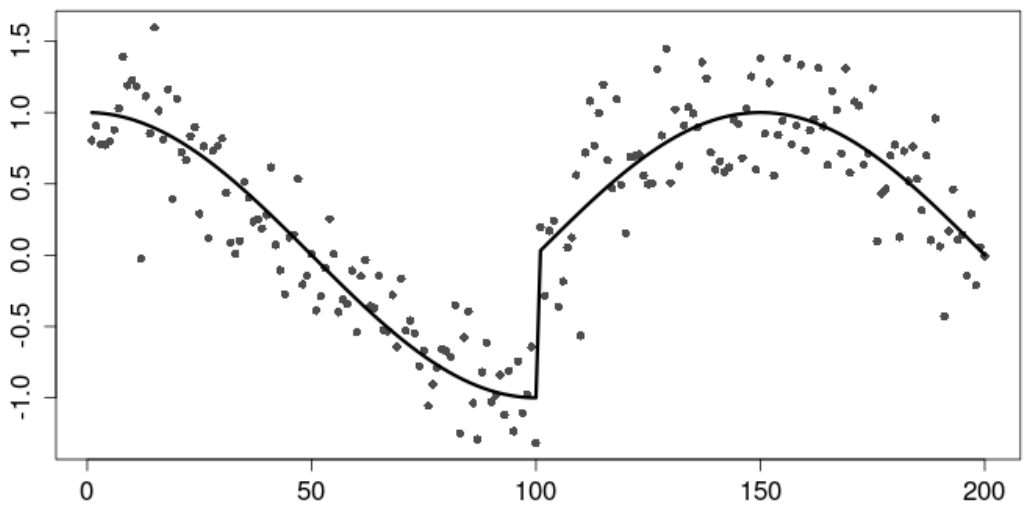}
\caption{The shifted cosine signal for $a=0.5$ (black line) and exemplary observations (grey points).}
\label{fig:signalRobustness}
\end{figure}

In this section we investigate how much our method is affected by a violation of the assumption that the signal can be decomposed into a piecewise constant plus a smooth function. To this end, we simulate $n=200$ observations from a signal with a single change-point plus a shifted cosine
\begin{equation*}
h(x) =  \begin{cases}
     \cos(2 \pi x) & \text{if } x < \frac{1}{2},\\
     1 + \cos(2 \pi x + a \pi) - (\cos((1 + a) \pi) - \cos(\pi)) & \text{otherwise},
   \end{cases}
\end{equation*}
for $a\in\{0, 0.1, 0.25, 0.5\}$. This signal has a single change-point of size $1$ at $1/2$ and the first derivative has a change of size $- 2\pi\sin((1+a) \pi)$ at the same location. Hence, this signal satisfies our model assumptions only if $a=0$ and the violation gets more severe with increasing $a\in (0,1/2]$. Subtraction of $\cos((1 + a) \pi) - \cos(\pi)$ ensures that size of the change is always $1$. The standard deviation is set to be $0.3$. An example for $a=0.5$ is presented in Figure \ref{fig:signalRobustness}. Simulation results are given in Tables \ref{tab:aMSErobustness} and \ref{tab:CProbustness}. Note that this setting is designed among other considerations such that the default bandwidth of $\JIC$ is a good choice to allow a comparison.

\begin{table}[ht]
\centering
\begin{tabular}{lcccc}
  \hline
Method & $a = 0$ & $a = 0.1$ & $a = 0.25$ & $a = 0.5$ \\ 
  \hline
$\pcs$ & \textbf{0.008832} & \textbf{0.008799} & \textbf{0.009323} & \textbf{0.01005} \\ 
  $\JIC$ & 0.01888 & 0.02113 & 0.02532 & 0.02614 \\ 
  $\PELT$ & 0.0377 & 0.03834 & 0.03676 & 0.03244 \\ 
   \hline
\end{tabular}
\caption{Averaged mean squared errors for the shifted cosine signal and various $a$.} 
\label{tab:aMSErobustness}
\end{table}

\begin{table}[ht]
\centering
\begin{tabular}{clccccccc}
  \hline
a & Method & $\hat{K} - K$ & $-1$ & $0$ & $1$ & $2$ & $> 2$ & \% detected \\ 
  \hline
0 & $\pcs$ & 1.027 &     9 & 27.95 & 39.98 & 11.81 & 11.26 &    90 \\ 
  0 & $\JIC$ & -0.1012 & 15.82 & 78.79 &  5.09 &  0.29 &  0.01 & 83.74 \\ 
  0 & $\COPS$ & 0.0802 &     0 & \textbf{94.09} &   3.8 &  2.11 &     0 & 99.89 \\ 
  0 & $\SOPS$ & -0.9995 & 99.99 &     0 &     0 &     0 &  0.01 &  0.01 \\ 
   \hline
0.1 & $\pcs$ & 1.051 &  8.72 & 23.75 & 45.73 & 11.09 & 10.71 & 90.46 \\ 
  0.1 & $\JIC$ & -0.0771 & 14.12 & 79.88 &  5.63 &  0.34 &  0.03 & 85.55 \\ 
  0.1 & $\COPS$ & 0.1102 &     0 & \textbf{92.05} &   4.9 &  3.03 &  0.02 & 99.86 \\ 
  0.1 & $\SOPS$ & -0.9995 & 99.99 &     0 &     0 &     0 &  0.01 &  0.01 \\ 
   \hline
0.25 & $\pcs$ & 1.084 &  9.33 & 21.22 & 47.84 &  9.99 & 11.62 & 89.84 \\ 
  0.25 & $\JIC$ & -0.0327 & 13.56 & 77.01 &  8.65 &  0.72 &  0.06 & 86.07 \\ 
  0.25 & $\COPS$ & 0.1276 &     0 & \textbf{90.83} &  5.62 &  3.51 &  0.04 & 99.71 \\ 
  0.25 & $\SOPS$ & -0.996 & 99.92 &     0 &     0 &     0 &  0.08 &  0.08 \\ 
   \hline
0.5 & $\pcs$ & 1.103 & 10.17 & 24.17 & 41.56 & 11.26 & 12.84 & 88.83 \\ 
  0.5 & $\JIC$ & 0.0004 & 11.29 & 78.44 &  9.27 &  0.94 &  0.06 & 88.23 \\ 
  0.5 & $\COPS$ & 0.1158 &     0 & \textbf{91.17} &  6.14 &  2.63 &  0.06 & 99.66 \\ 
  0.5 & $\SOPS$ & -0.997 & 99.94 &     0 &     0 &     0 &  0.06 &  0.05 \\ 
   \hline
\end{tabular}
\caption{Summary results about the detection of change-points for the shifted cosine signal and various $a$.} 
\label{tab:CProbustness}
\end{table}

Tables \ref{tab:aMSErobustness} and \ref{tab:CProbustness} show that $\PCpluS$ is remarkably robust against such a violation of its assumptions. Not only does $\PCpluS$ have the smallest MSE in all cases, but its performance deteriorates only slightly as $a$ increases, in contrast to $\JIC$ for example, which is designed for such cases.

All in all, we find that $\PCpluS$ performs well and with respect to the MSE in particular, and it is outperformed only on rare occasions. It also shows good change-point detection results, but has a slight tendency to overestimate the number of change-points. Moreover, it is rather robust to a violation of the assumption that the signal can be decomposed into a piecewise constant plus a smooth function.

\section{Application: Genome sequencing}\label{sec:application}
Change-point regression is commonly used in the analysis of genome sequencing data for detecting copy number variations, see e.g. \citet{olshen2004circular, zhang2007modified, futschik2014multiscale, hocking2020constrained}. Copy number variations are amplifications or deletions of genetic material, which can range from a few base pairs to a whole chromosome. Detecting copy number variations is of high importance as they can be linked to diseases such as cancer. To this end, full genome sequencing techniques, such as array-based comparative genomic hybridization (CGH) and next-generation sequencing (NGS), are used to measure the copy number at several thousands genome locations simultaneously. The resulting data are normalized $\log_2$ ratios of recoded samples of short DNA sequences of the genome and a reference genome. Hence, values around zero indicate the normal two copies, while in an ideal situation $-1$ stands for a single deletion and $\log_2(3)\approx 0.58$ for three copies. Details can be found in the afore mentioned publications.

However, artefacts occur frequently. The two most important ones are mappability and GC content bias. Mappability bias means that some reads cannot matched uniquely to a certain region in the genome; GC content bias means that regions with medium GC content (the amount of the chemical bases guanine G and cytosine C) are more likely mapped than others. Though some bias corrections are available, some artefacts still remain, either because the corrections do not work perfectly or because further unknown biases occur. We refer the reader to  \citet{liu2013computational} for further background. Hence, the classical model of a piecewise constant signal plus centred random errors is violated. 

In the following we use $\PCpluS$ to analyse the Coriell cell line GM03576 in the array CGH data set from \citet{snijders2001assembly}. It can almost be seen as a `gold standard' for genome sequencing, mainly because for this dataset the truth is already explored by spectral karyotyping. In cell line GM03576, trisomies for chromosomes 2 and 21 are known.

\begin{figure}[!htb]
\includegraphics[width = 0.9\textwidth]{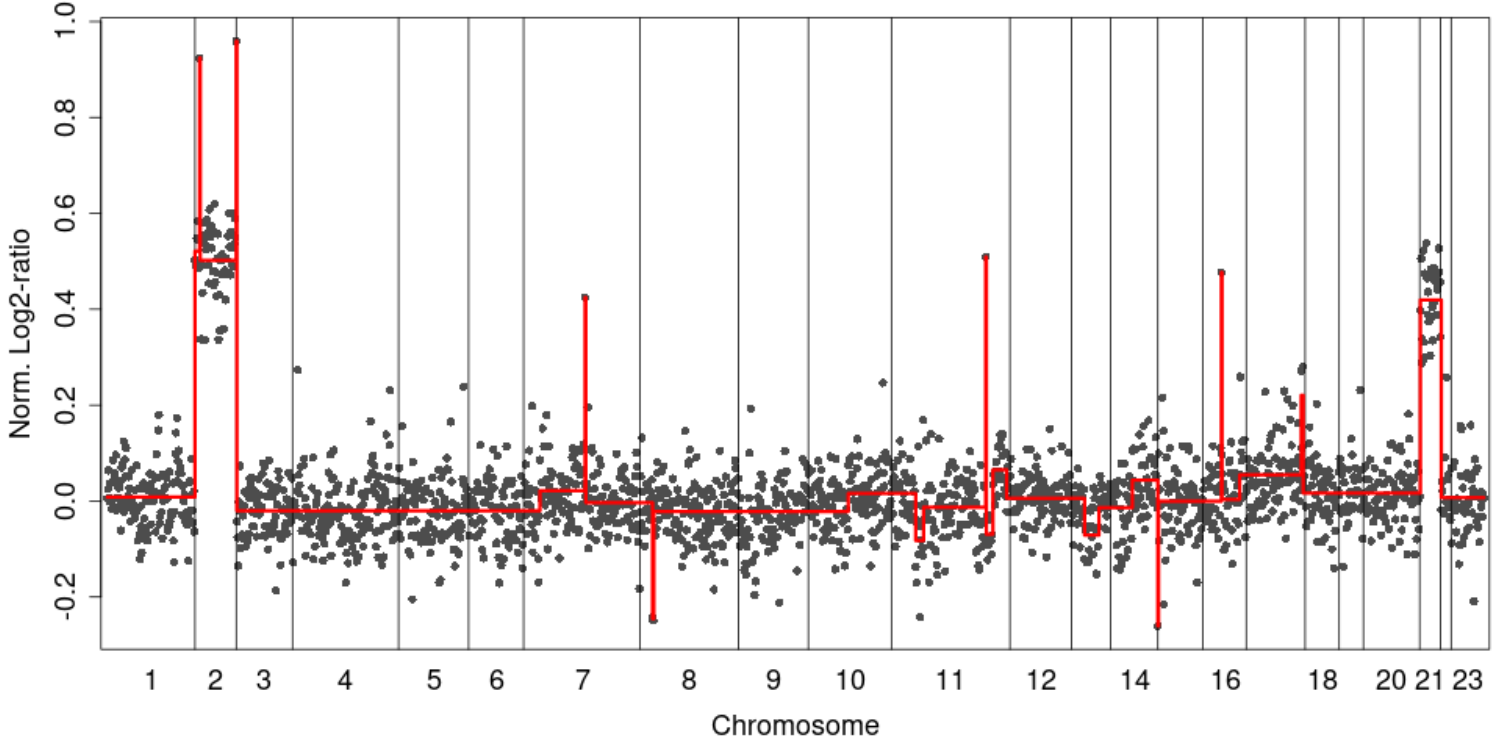} 
\caption{Cell line GM03576 analysed by $\PELT$. It detects both trisomies and several single point outliers, but also several small changes, which are most likely false positives.}
\label{fig:GM03576pelt}
\end{figure}

In Figure \ref{fig:GM03576pelt} we find that $\PELT$ with the $\operatorname{SIC}$-penalty, which we use as an example for a piecewise constant change-point regression approach, detects both trisomies, but also single point outliers and several small changes. The positive outliers could potentially be linked to additional amplifications, but all other detections are most likely false positives.

\begin{figure}[!htb]
\includegraphics[width = 0.9\textwidth]{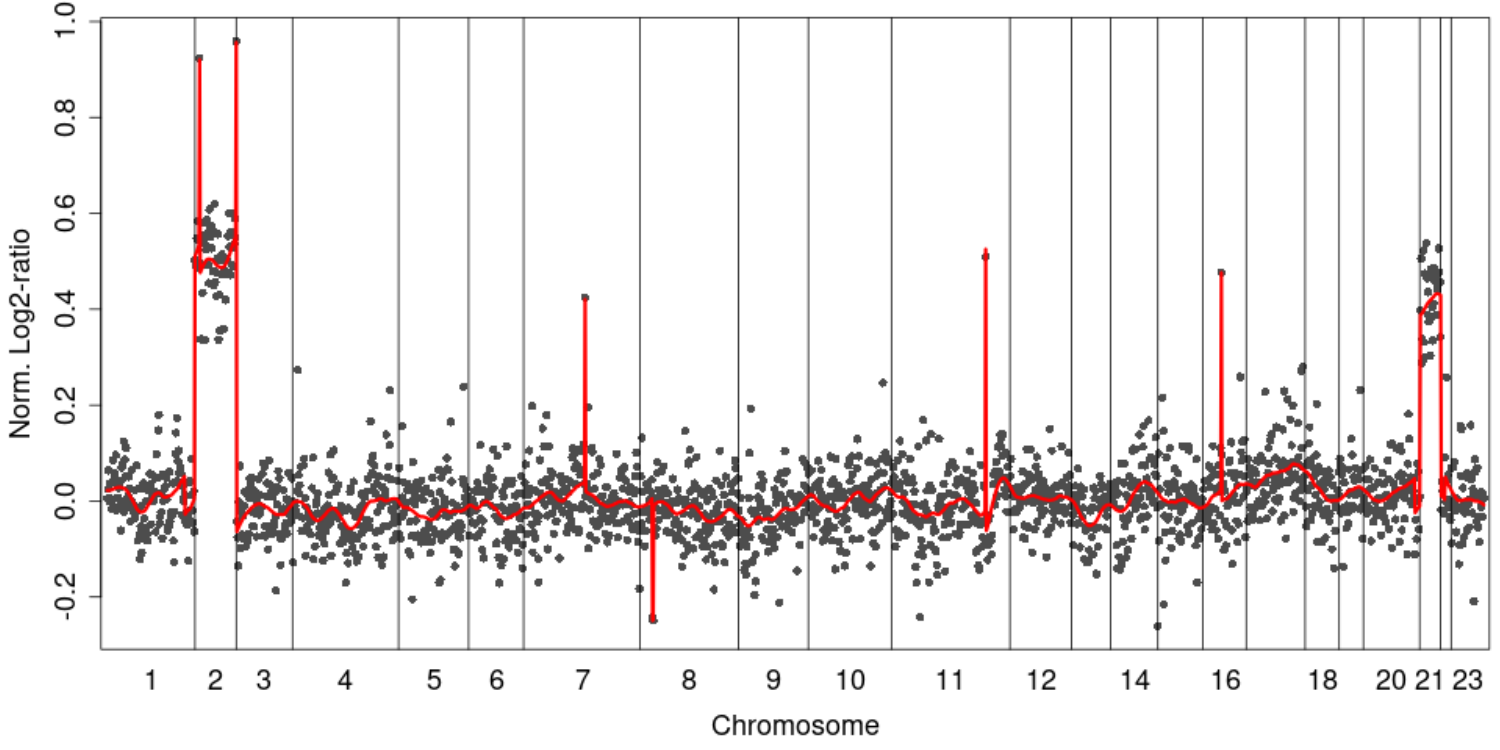} 
\caption{Cell line GM03576 analysed by cross-validated $\PCpluS$. It also detects both trisomies and the positive outliers, but only one negative outlier and three false positives next to existing events.}
\label{fig:GM03576pcplus}
\end{figure}

Figure \ref{fig:GM03576pcplus} shows the corresponding fit using cross-validated $\PCpluS$. 
It detects both trisomies and the positive outliers, too, but only one negative outlier and three other false positives next to the true trisomies. In Appendix \ref{appendix:furtherApplication} we show that the additional false positives can be avoided by choosing slightly different tuning parameters. Overall then, $\PCpluS$ successfully detects all critical change-points, but almost no additional false positives.

\section{Discussion}\label{sec:discussion}
We proposed $\PCpluS$, a combination of the fused Lasso and linear smoother, which is to the best of our knowledge the first approach that explicitly uses the decomposition into a piecewise constant function and a smooth function when detecting change-points. We demonstrated its good performance in simulations and on real data and detailed an efficient procedure for its computation when kernel smoothing is used.
The general approach can also be adapted to many more complicated settings quite easily, such as the settings of multivariate data and filtered data.

\subsection{Multivariate data}\label{sec:multivariate}
In this section we outline how $\PCpluS$ can be extended to multivariate data. This is for instance of interest in genome sequencing when multiple patients with the same form of cancer are studied. In such situations, it is reasonable to assume that all patients have changes at the same locations in the genome; for more details see \citet{bleakley2011group}. This suggests model for the response $Y_{ij}$ at location $i$ of the $j$th patient of the form
\begin{equation*}
Y_{ij}=f^0_j(i/n) + g^0_j(i/n) + \varepsilon_{ij},\ i=1,\ldots,n,\ j=1,\ldots,p,
\end{equation*}
where $f^0_j$ are piecewise constant signals with changes at the same locations and $g^0_j$ are individual smooth functions: here we assume that the smooth artefacts are not shared among the patients. Instead of basing our initial estimation on the fused Lasso, we can instead use the group fused Lasso \citep{bleakley2011group}:
\begin{equation*}
\hat{f}:=\argmin_{f \in \R^{n\times p}} \left\{\sum_{j=1}^p \| (I-S^{(j)}) (Y_{j} - f_{j}) \|_2^2  + \lambda \sum_{i=1}^{n - 1} \left(\sum_{j=1}^p (f_{(i+1)j} - f_{ij})^2\right)^{1/2} \right\}.
\end{equation*}
Here the $S^{(j)}$ are smoothing matrices corresponding to each patient and $Y_j$ and $f_j$ are the $j$th columns of the matrices $Y$ and $f$ respectively.

\subsection{Filtered data}\label{sec:filter}
Finally, we outline how $\PCpluS$ can be adapted to filtered data as are common in the analysis of ion channel recordings \citep{neher1976single, sakmann2013single}, which are experiments to measure the conductance of a single ion channel over time. Ion channels are poreforming proteins in the cell membrane that allows ions to pass the membrane since they are needed for many vital processes in the cells. Channels regulate the ion flow by opening and closing over time. Studying those gating dynamics is of interest for instance in medicine or biochemistry. The conductance of an ion channel can be modelled as piecewise constant, but smooth artefacts are common, for instance caused by the electronic, building vibrations or small holes in the membrane. The recordings are filtered by lowpass filters, often Bessel filters, which are integrated in the hardware and hence known. The time continuous filter can be discretized at small errors. For further details we refer to \citet{pein2018fully} and the references therein. Writing $F \in \R^{n \times n}$ we obtain  the model
\begin{equation*}
Y_i := (F  f^*)_i + (F g^*)_i  + \varepsilon_i, \quad i = 1,\ldots,n.
\end{equation*}
We note that if in other experiments the filter is unknown, it can be pre-estimated using the techniques in \citet{tecuapetla2017autocovariance}. Note that $F g^*$ is still a vector of evaluations of a smooth function so adding filtering to $\PCpluS$ to estimate the piecewise constant signal is straightforward. The initial step \eqref{eq:modifiedFusedLasso} is simply replaced by
\begin{equation}\label{eq:estFilter}
\hat{f}:=\argmin_{f \in \R^{n}} \| (I - S) (Y - F  f)\|_2^2 + \lambda \| f \|_{\operatorname{TV}}.
\end{equation}
The outlined extensions demonstrate that $\PCpluS$ is a flexible approach that can often be adapted to more involved settings easily. In many cases an extension of $\PCpluS$ only requires that one is able to adapt the fused Lasso and the smoothing separately to the new problem.

%
%
%
%
%


\bibliographystyle{apalike}
\bibliography{Literature}

\begin{thebibliography}{}

\bibitem[Abramovich et~al., 2007]{abramovich2007estimation}
Abramovich, F., Antoniadis, A., and Pensky, M. (2007).
\newblock Estimation of piecewise-smooth functions by amalgamated bridge
  regression splines.
\newblock {\em Sankhya}, pages 1--27.

\bibitem[Antoniadis and Gijbels, 2002]{AntoniadisGijbels02}
Antoniadis, A. and Gijbels, I. (2002).
\newblock Detecting abrupt changes by wavelet methods.
\newblock {\em J. Nonparametr. Stat.}, 14(1-2):7--29.

\bibitem[Bai and Perron, 2003]{bai2003computation}
Bai, J. and Perron, P. (2003).
\newblock Computation and analysis of multiple structural change models.
\newblock {\em J. Appl. Econ.}, 18(1):1--22.

\bibitem[Bleakley and Vert, 2011]{bleakley2011group}
Bleakley, K. and Vert, J.-P. (2011).
\newblock The group fused lasso for multiple change-point detection.
\newblock {\em arXiv preprint arXiv:1106.4199}.

\bibitem[Bolorinos et~al., 2020]{bolorinos2020consumption}
Bolorinos, J., Ajami, N.~K., and Rajagopal, R. (2020).
\newblock Consumption change detection for urban planning: {M}onitoring and
  segmenting water customers during drought.
\newblock {\em Water Resour. Res.}, 56(3):e2019WR025812.

\bibitem[Denby, 1984]{denby1984smooth}
Denby, L. (1984).
\newblock Smooth regression functions.
\newblock {\em Diss. Abst. Int. Pt. B - Sci. \& Eng.}, 45(2).

\bibitem[Denison et~al., 1998]{Denisonetal98}
Denison, D., Mallick, B., and Smith, A. (1998).
\newblock Automatic {B}ayesian curve fitting.
\newblock {\em J. R. Stat. Soc. Series B Stat. Methodol.}, 60(2):333--350.

\bibitem[DiMatteo et~al., 2001]{Dimatteoetal01}
DiMatteo, I., Genovese, C.~R., and Kass, R.~E. (2001).
\newblock Bayesian curve-fitting with free-knot splines.
\newblock {\em Biometrika}, 88(4):1055--1071.

\bibitem[D’Angelo et~al., 2011]{d2011incipient}
D’Angelo, M., Palhares, R.~M., Takahashi, R., Loschi, R.~H., Baccarini, L.,
  and Caminhas, W. (2011).
\newblock Incipient fault detection in induction machine stator-winding using a
  fuzzy-bayesian change point detection approach.
\newblock {\em Appl. Soft. Comput.}, 11(1):179--192.

\bibitem[Eubank and Speckman, 1994]{eubank1994nonparametric}
Eubank, R.~L. and Speckman, P.~L. (1994).
\newblock Nonparametric estimation of functions with jump discontinuities.
\newblock {\em Lect. Notes Monogr. Ser.}, pages 130--144.

\bibitem[Fearnhead, 2005]{Fearnhead05}
Fearnhead, P. (2005).
\newblock Exact {B}ayesian curve fitting and signal segmentation.
\newblock {\em IEEE Trans. Signal Process.}, 53(6):2160--2166.

\bibitem[Frick et~al., 2014]{frick2014multiscale}
Frick, K., Munk, A., and Sieling, H. (2014).
\newblock Multiscale change point inference.
\newblock {\em J. R. Stat. Soc. Series B Stat. Methodol.}, 76(3):495--580.

\bibitem[Friedman et~al., 2007]{friedman2007pathwise}
Friedman, J., Hastie, T., H{\"o}fling, H., and Tibshirani, R. (2007).
\newblock Pathwise coordinate optimization.
\newblock {\em Ann. Appl. Stat.}, 1(2):302--332.

\bibitem[Friedman et~al., 2010]{friedman2010regularization}
Friedman, J.~H., Hastie, T., and Tibshirani, R. (2010).
\newblock Regularization paths for generalized linear models via coordinate
  descent.
\newblock {\em Journal of statistical software}, 33:1--22.

\bibitem[Futschik et~al., 2014]{futschik2014multiscale}
Futschik, A., Hotz, T., Munk, A., and Sieling, H. (2014).
\newblock Multiscale {DNA} partitioning: statistical evidence for segments.
\newblock {\em Bioinformatics}, 30(16):2255--2262.

\bibitem[Gijbels and Goderniaux, 2004]{GijbelsGoderniaux04}
Gijbels, I. and Goderniaux, A.-C. (2004).
\newblock Bandwidth selection for changepoint estimation in nonparametric
  regression.
\newblock {\em Technometrics}, 46(1):76--86.

\bibitem[Gijbels et~al., 2004]{Gjbelsetal04}
Gijbels, I., Hall, P., and Kneip, A. (2004).
\newblock Interval and band estimation for curves with jumps.
\newblock {\em J. Appl. Probab.}, 41(A):65--79.

\bibitem[Gijbels et~al., 2007]{Gijbelsetal07}
Gijbels, I., Lambert, A., and Qiu, P. (2007).
\newblock Jump-preserving regression and smoothing using local linear fitting:
  a compromise.
\newblock {\em Ann. Inst. Stat. Math.}, 59(2):235--272.

\bibitem[Harchaoui et~al., 2009]{harchaoui2009regularized}
Harchaoui, Z., Vallet, F., Lung-Yut-Fong, A., and Capp{\'e}, O. (2009).
\newblock A regularized kernel-based approach to unsupervised audio
  segmentation.
\newblock In {\em 2009 IEEE International Conference on Acoustics, Speech and
  Signal Processing}, pages 1665--1668. IEEE.

\bibitem[Hocking et~al., 2020]{hocking2020constrained}
Hocking, T.~D., Rigaill, G., Fearnhead, P., and Bourque, G. (2020).
\newblock Constrained dynamic programming and supervised penalty learning
  algorithms for peak detection in genomic data.
\newblock {\em J. Mach. Learn. Res.}, 21:1--28.

\bibitem[Hotz et~al., 2013]{hotz2013idealizing}
Hotz, T., Sch{\"u}tte, O.~M., Sieling, H., Polupanow, T., Diederichsen, U.,
  Steinem, C., and Munk, A. (2013).
\newblock Idealizing ion channel recordings by a jump segmentation
  multiresolution filter.
\newblock {\em IEEE Trans. Nanobioscience}, 12(4):376--386.

\bibitem[Huang and Qiu, 2010]{HuangQui10}
Huang, X. and Qiu, P. (2010).
\newblock Blind deconvolution for jump-preserving curve estimation.
\newblock {\em Math. Probl. Eng.}, 2010.

\bibitem[Killick et~al., 2012]{killick2012optimal}
Killick, R., Fearnhead, P., and Eckley, I.~A. (2012).
\newblock Optimal detection of changepoints with a linear computational cost.
\newblock {\em J. Am. Stat. Assoc.}, 107(500):1590--1598.

\bibitem[Kim et~al., 2005]{kim2005structural}
Kim, C.-J., Morley, J.~C., and Nelson, C.~R. (2005).
\newblock The structural break in the equity premium.
\newblock {\em J. Bus. Econ. Stat.}, 23(2):181--191.

\bibitem[Koo, 1997]{Koo97}
Koo, J.-Y. (1997).
\newblock Spline estimation of discontinuous regression functions.
\newblock {\em J. Comput. Graph. Stat.}, 6(3):266--284.

\bibitem[L., 2018]{Lee18}
L., Y. (2018).
\newblock {\em Bayesian curve fitting for discontinuous functions using
  overcomplete system with multiple kernels}.
\newblock PhD thesis, Seoul National University.

\bibitem[Lam et~al., 2016]{Lametal16}
Lam, B.~S., Carisa, K., Choy, S.-K., and Leung, J.~K. (2016).
\newblock Jump point detection using empirical mode decomposition.
\newblock {\em Land Use Policy}, 58:1--8.

\bibitem[Lee and Jhong, 2021]{lee2021change}
Lee, E.-J. and Jhong, J.-H. (2021).
\newblock Change point detection using penalized multidegree splines.
\newblock {\em Axioms}, 10(04):331.

\bibitem[Lee, 2002]{Lee02}
Lee, T.~C. (2002).
\newblock Automatic smoothing for discontinuous regression functions.
\newblock {\em Stat. Sin.}, pages 823--842.

\bibitem[Lin et~al., 2017]{lin2017sharp}
Lin, K., Sharpnack, J.~L., Rinaldo, A., and Tibshirani, R.~J. (2017).
\newblock A sharp error analysis for the fused lasso, with application to
  approximate changepoint screening.
\newblock In {\em Advances in Neural Information Processing Systems}, pages
  6884--6893.

\bibitem[Liu et~al., 2013]{liu2013computational}
Liu, B., Morrison, C.~D., Johnson, C.~S., Trump, D.~L., Qin, M., Conroy, J.~C.,
  Wang, J., and Liu, S. (2013).
\newblock Computational methods for detecting copy number variations in cancer
  genome using next generation sequencing: principles and challenges.
\newblock {\em Oncotarget}, 4(11):1868.

\bibitem[Liu et~al., 2018]{Liuetal18}
Liu, G.-X., Wang, M.-M., Du, X.-L., Lin, J.-G., and Gao, Q.-B. (2018).
\newblock Jump-detection and curve estimation methods for discontinuous
  regression functions based on the piecewise {B}-spline function.
\newblock {\em Comm. Statist. Theory Methods}, 47(23):5729--5749.

\bibitem[Loader, 1996]{Loader96}
Loader, C.~R. (1996).
\newblock Change point estimation using nonparametric regression.
\newblock {\em Ann. Stat.}, 24(4):1667--1678.

\bibitem[Lu, 2023]{lu2023simultaneous}
Lu, Z. (2023).
\newblock {\em Simultaneous Change-point Detection and Curve Estimation for
  Single and Multiple Sequential Data}.
\newblock PhD thesis, The University of Arizona.

\bibitem[Lung-Yut-Fong et~al., 2012]{lung2012distributed}
Lung-Yut-Fong, A., L{\'e}vy-Leduc, C., and Capp{\'e}, O. (2012).
\newblock Distributed detection/localization of change-points in
  high-dimensional network traffic data.
\newblock {\em Stat. Comput.}, 22(2):485--496.

\bibitem[Mboup et~al., 2008]{Mboupetal08}
Mboup, M., Join, C., and Fliess, M. (2008).
\newblock An on-line change-point detection method.
\newblock In {\em Control and Automation, 2008 16th Mediterranean Conference},
  pages 1290--1295. IEEE.

\bibitem[Miyata and Shen, 2003]{MiyataShen03}
Miyata, S. and Shen, X. (2003).
\newblock Adaptive free-knot splines.
\newblock {\em J. Comput. Graph. Stat.}, 12(1):197--213.

\bibitem[Moreno et~al., 2013]{Morenoetal13}
Moreno, E., Javier~G., F., and Garc{\'\i}a-Ferrer, A. (2013).
\newblock A consistent on-line {B}ayesian procedure for detecting change
  points.
\newblock {\em Environmetrics}, 24(5):342--356.

\bibitem[M{\"u}ller, 1992]{Mueller92}
M{\"u}ller, H.-G. (1992).
\newblock Change-points in nonparametric regression analysis.
\newblock {\em Ann. Stat.}, pages 737--761.

\bibitem[Neher and Sakmann, 1976]{neher1976single}
Neher, E. and Sakmann, B. (1976).
\newblock Single-channel currents recorded from membrane of denervated frog
  muscle fibres.
\newblock {\em Nature}, 260(5554):799--802.

\bibitem[Niu and Zhang, 2012]{niu2012screening}
Niu, Y.~S. and Zhang, H. (2012).
\newblock The screening and ranking algorithm to detect {DNA} copy number
  variations.
\newblock {\em Ann. Appl. Stat.}, 6(3):1306.

\bibitem[Norouzirad et~al., 2022]{norouzirad2022differenced}
Norouzirad, M., Arashi, M., and Roozbeh, M. (2022).
\newblock Differenced-based double shrinking in partial linear models.
\newblock {\em Journal of Data Science and Modeling}, 1(1):21--32.

\bibitem[Ogden, 1999]{Ogden99}
Ogden, R.~T. (1999).
\newblock Wavelets in the {B}ayesian change-point analysis.
\newblock In {\em Wavelet Applications in Signal and Image Processing VII},
  volume 3813, pages 245--251. International Society for Optics and Photonics.

\bibitem[Olshen et~al., 2004]{olshen2004circular}
Olshen, A.~B., Venkatraman, E., Lucito, R., and Wigler, M. (2004).
\newblock Circular binary segmentation for the analysis of array-based {DNA}
  copy number data.
\newblock {\em Biostatistics}, 5(4):557--572.

\bibitem[Pein, 2025]{PCpluS}
Pein, F. (2025).
\newblock {\em {PCpluS}: Piecewise Constant Plus Smooth Regression}.
\newblock R package version 1.0.

\bibitem[Pein and Shah, 2021]{pein2021crossvalidation}
Pein, F. and Shah, R.~D. (2021).
\newblock Cross-validation for change-point regression: pitfalls and solutions.
\newblock {\em arxiv:2112.03220}.

\bibitem[Pein et~al., 2017]{pein2017heterogeneous}
Pein, F., Sieling, H., and Munk, A. (2017).
\newblock Heterogeneous change point inference.
\newblock {\em J. R. Stat. Soc. Series B Stat. Methodol.}, 79(4):1207--1227.

\bibitem[Pein et~al., 2018]{pein2018fully}
Pein, F., Tecuapetla-G{\'o}mez, I., Sch{\"u}tte, O.~M., Steinem, C., and Munk,
  A. (2018).
\newblock Fully automatic multiresolution idealization for filtered ion channel
  recordings: flickering event detection.
\newblock {\em IEEE Trans. Nanobioscience}, 17(3):300--320.

\bibitem[Punskaya et~al., 2002]{Punskayaetal02}
Punskaya, E., Andrieu, C., Doucet, A., and Fitzgerald, W.~J. (2002).
\newblock Bayesian curve fitting using {MCMC} with applications to signal
  segmentation.
\newblock {\em IEEE Trans. Signal Process.}, 50(3):747--758.

\bibitem[Qian and Jia, 2016]{qian2016stepwise}
Qian, J. and Jia, J. (2016).
\newblock On stepwise pattern recovery of the fused {L}asso.
\newblock {\em Comput. Stat. Data Anal.}, 94:221--237.

\bibitem[Qiu, 1994]{Qiu94}
Qiu, P. (1994).
\newblock Estimation of the number of jumps of the jump regression functions.
\newblock {\em Comm. Statist. Theory Methods}, 23(8):2141--2155.

\bibitem[Qiu, 2003]{Qiu03}
Qiu, P. (2003).
\newblock A jump-preserving curve fitting procedure based on local
  piecewise-linear kernel estimation.
\newblock {\em J. Nonparametr. Stat.}, 15(4-5):437--453.

\bibitem[Qiu, 2005]{qiu2005image}
Qiu, P. (2005).
\newblock {\em Image processing and jump regression analysis}, volume 599.
\newblock John Wiley \& Sons.

\bibitem[Qiu and Yandell, 1998]{QiuYandell02}
Qiu, P. and Yandell, B. (1998).
\newblock Local polynomial jump-detection algorithm in nonparametric
  regression.
\newblock {\em Technometrics}, 40(2):141--152.

\bibitem[Raimondo, 1998]{Raimondo98}
Raimondo, M. (1998).
\newblock Minimax estimation of sharp change points.
\newblock {\em Ann. Stat.}, pages 1379--1397.

\bibitem[Reeves et~al., 2007]{reeves2007review}
Reeves, J., Chen, J., Wang, X.~L., Lund, R., and Lu, Q. (2007).
\newblock A review and comparison of changepoint detection techniques for
  climate data.
\newblock {\em J. Appl. Meteorol. Climatol.}, 46(6):900--915.

\bibitem[Sakmann, 2013]{sakmann2013single}
Sakmann, B. (2013).
\newblock {\em Single-channel recording}.
\newblock Springer Science \& Business Media.

\bibitem[Snijders et~al., 2001]{snijders2001assembly}
Snijders, A.~M., Nowak, N., Segraves, R., Blackwood, S., Brown, N., Conroy, J.,
  Hamilton, G., Hindle, A.~K., Huey, B., and Kimura, K. (2001).
\newblock Assembly of microarrays for genome-wide measurement of {DNA} copy
  number.
\newblock {\em Nat. Genet.}, 29(3):263--264.

\bibitem[Speckman, 1988]{speckman1988kernel}
Speckman, P. (1988).
\newblock Kernel smoothing in partial linear models.
\newblock {\em J. R. Stat. Soc. Series B Stat. Methodol.}, 50(3):413--436.

\bibitem[Spokoiny, 1998]{Spokoiny98}
Spokoiny, V.~G. (1998).
\newblock Estimation of a function with discontinuities via local polynomial
  fit with an adaptive window choice.
\newblock {\em Ann. Stat.}, 26(4):1356--1378.

\bibitem[Tecuapetla-G{\'o}mez and Munk, 2017]{tecuapetla2017autocovariance}
Tecuapetla-G{\'o}mez, I. and Munk, A. (2017).
\newblock Autocovariance estimation in regression with a discontinuous signal
  and m-dependent errors: {A} difference-based approach.
\newblock {\em Scand. J. Stat.}, 44(2):346--368.

\bibitem[Tibshirani et~al., 2012]{tibshirani2012strong}
Tibshirani, R., Bien, J., Friedman, J., Hastie, T., Simon, N., Taylor, J., and
  Tibshirani, R.~J. (2012).
\newblock Strong rules for discarding predictors in lasso-type problems.
\newblock {\em Journal of the Royal Statistical Society Series B: Statistical
  Methodology}, 74(2):245--266.

\bibitem[Tibshirani et~al., 2005]{tibshirani2005sparsity}
Tibshirani, R., Saunders, M., Rosset, S., Zhu, J., and Knight, K. (2005).
\newblock Sparsity and smoothness via the fused lasso.
\newblock {\em J. R. Stat. Soc. Series B Stat. Methodol.}, 67(1):91--108.

\bibitem[Wang et~al., 2022]{wang2022data}
Wang, G., Zou, C., and Qiu, P. (2022).
\newblock Data-driven determination of the number of jumps in regression
  curves.
\newblock {\em Technometrics}, 64(3):312--322.

\bibitem[Wang, 1995]{Wang95}
Wang, Y. (1995).
\newblock Jump and sharp cusp detection by wavelets.
\newblock {\em Biometrika}, 82(2):385--397.

\bibitem[Wu and Zhao, 2007]{WuZhao07}
Wu, W.~B. and Zhao, Z. (2007).
\newblock Inference of trends in time series.
\newblock {\em J. R. Stat. Soc. Series B Stat. Methodol.}, 69(3):391--410.

\bibitem[Xia and Qiu, 2015]{xia2015jump}
Xia, Z. and Qiu, P. (2015).
\newblock Jump information criterion for statistical inference in estimating
  discontinuous curves.
\newblock {\em Biometrika}, 102(2):397--408.

\bibitem[Yang and Song, 2014]{YangSong14}
Yang, Y. and Song, Q. (2014).
\newblock Jump detection in time series nonparametric regression models: a
  polynomial spline approach.
\newblock {\em Ann. Inst. Stat. Math.}, 66(2):325--344.

\bibitem[Younes et~al., 2014]{younes2014inferring}
Younes, L., Albert, M., Miller, M.~I., and Team, B.~R. (2014).
\newblock Inferring changepoint times of medial temporal lobe morphometric
  change in preclinical alzheimer's disease.
\newblock {\em Neuroimage Clin.}, 5:178--187.

\bibitem[Zhang and Siegmund, 2007]{zhang2007modified}
Zhang, N.~R. and Siegmund, D.~O. (2007).
\newblock A modified {B}ayes information criterion with applications to the
  analysis of comparative genomic hybridization data.
\newblock {\em Biometrics}, 63(1):22--32.

\bibitem[Zhang, 2016]{Zhang16}
Zhang, T. (2016).
\newblock Testing for jumps in the presence of smooth changes in trends of
  nonstationary time series.
\newblock {\em Electron. J. Stat.}, 10(1):706--735.

\end{thebibliography}

\appendix

\newpage

\section{Proofs}\label{appendix:proofs}

\begin{proof}[Proof of Proposition~\ref{prop:X_tilde}]
	Recalling that $X_{lj} = \ind_{\{l>j\}}$, we see that
	\[
	(KX)_{ij} = \sum_{l=j + 1}^n K_{il} = \ind_{\{j+1 \leq i + L\}} \sum_{l=(j + 1) \vee (i-L)}^{n \wedge (i+L)} k_{l-i} = \ind_{\{j -i \leq L-1\}} \sum_{l=(j+1-i) \vee -L}^{(n-i) \wedge L} k_{l}.
	\]
	We thus see that by first computing the cumulative sums of the sequence $k_{-L}, k_{-L + 1} \ldots, k_L$, we can obtain any element $(KX)_{ij}$ in constant time. Since we may then also obtain any element of $D$ \eqref{eq:D_def} in constant time, $(SX)_{ij} = D_{ii} (KX)_{ij}$ and hence $\tilde{X}_{ij}$  may be obtained in constant time.

	Recalling that
	\[
	D_{ii} = \left( \sum_{l=(1-i) \vee -L}^{(n-i) \wedge L} k_l \right)^{-1},
	\]
	we also see that $D_{ii} (KX)_{ij} = 0$ if $j < i-L$. But when $i - L > j$, $X_{ij}=1$, so putting things together, we see that $(SX)_{ij} = 0$ if $|i-j| > L$.
	
	If $|i-j| \leq L$, and $L \leq i \leq n-L$, then from the above,
	\[
	(KX)_{ij} = \sum_{l=j-i}^L k_l \qquad \text{and} \qquad D_{ii} = \left(\sum_{l=-L}^L k_l \right)^{-1}.
	\]
	Thus for such $i,j$, the value of $(SX)_{ij}$ depends only on $j-i$, and hence the $(n-2L) \times (n-1)$ submatrix of $\tilde{X}$ formed through excluding the first $L$ and last $L$ rows is Toeplitz as required.
\end{proof}

\begin{proof}[Proof of Proposition~\ref{proposition:tildeXtildeX}]
We have from Proposition~\ref{prop:X_tilde} that
\begin{equation*}
	\left(\tilde{X}^\top \tilde{X}\right)_{ij}
	= \sum_{l = 1}^{n} \tilde{X}_{li} \tilde{X}_{lj}
	= \ind_{\{\min(i + L, j + L) >  \max(i - L, j - L)\}} \sum_{l = \max(i - L, j - L, 1)}^{\min(i + L, j + L, n)} \tilde{X}_{li} \tilde{X}_{lj}.
\end{equation*}
Thus if $|i-j| > 2L$, then $\big(\tilde{X}^\top \tilde{X}\big)_{ij} = 0$.

For the second point, Proposition~\ref{prop:X_tilde} implies that $\tilde{X}_{ij} = \tilde{X}_{(i + k)(j + k)}$ if $L < i \leq n - L$, $k$ is such that $L < i + k \leq n - L$ and $1 \leq j + k < n - 1$. Thus from the above, for $2L < i < n-2L$ and $k \leq (n - 2L - i) \wedge (n-j)$, we have
\[
\left(\tilde{X}^\top \tilde{X}\right)_{ij} =  \sum_{l = \max(i - L, j - L, 1)}^{\min(i + L, j + L, n)} \tilde{X}_{(l+k)(i+k)} \tilde{X}_{(l+k)(j+k)} = \left(\tilde{X}^\top \tilde{X}\right)_{(i+k)(j+k)}.\qedhere
\]
\end{proof}

\section{Pseudocode}\label{appendix:pseudocode}

\begin{algorithm}[ht]
\caption{Overview of the steps of $\PCpluS$.}
\label{alg:pcplus}
\begin{algorithmic}[1]
\Require
Regularisation parameter $\lambda >0$ and bandwidth $h$ (typically grids of values from which the final pair can be selected by cross-validation, see Section~\ref{sec:crossvalidation}).
\State Fused Lasso combined with kernel smoothing:
\begin{equation*}
\hat{f} := \argmin_{f \in \R^n,\ f_1 = 0} \| (I - S) (Y - f)\|_2^2 + \lambda \| f \|_{\operatorname{TV}},
\end{equation*}
where $S$ is the kernel smoothing matrix.
\State \begin{equation*}
\hat{g}:=\hat{g}(\hat{f})=S (Y - \hat{f})
\end{equation*}
\State Post-processing step 1: Piecewise-constant fit to $Y - \hat{g}$:
\begin{equation*}
\tilde{f} := \argmin_{f \in \R^n} \| (Y - \hat{g}) - f\|_2^2 + \lambda_{\PELT} \| f \|_0
\end{equation*}
\State Post-processing step 2: Estimate $f^0$ without penalty, but with given change-point set $\hat{J}:=\{n\hat{\tau}_1,\ldots,n\hat{\tau}_{\hat{K}}\}$, where $\hat{\tau}_1,\ldots,\hat{\tau}_{\hat{K}}$ are the change-points of $\tilde{f}$:
\begin{equation*}
\hat{f}^{\mathrm{PC}+} := \argmin_{\substack{f \in \R^n,\ f_1 = 0,\\ f_i\neq f_{i+1}\ \Leftrightarrow\ i \in \hat{J}}} \|(I - S) (Y - \tilde{f})\|_2^2
\end{equation*}
\begin{equation*}
\hat{g}^{\mathrm{PC}+}:=S (Y - \hat{f}^{\mathrm{PC}+})
\end{equation*}
\Return $(\hat{f}^{\mathrm{PC}+}, \hat{g}^{\mathrm{PC}+})$
\end{algorithmic}
\end{algorithm}

\section{Further results}
\subsection{Simulation}\label{sec:appendixSimulationResults}
Tables \ref{tab:aMSEn256a6}--\ref{tab:CPn256a4} report further simulation results that are discussed in the main paper.

\begin{table}[ht]
\centering
\begin{tabular}{lcccc}
  \hline
Method & blocks & burt & cosine & heavisine \\ 
  \hline
$\pcs$ & \textbf{0.01118} & 0.5144 & \textbf{0.117} & 0.0456 \\ 
  $\JIC$ & 1.841 & 23.72 & 0.3047 & 1.629 \\ 
    $\operatorname{ABS1}$ & 0.47 & 0.43 & 0.25 & 0.21\\
  $\PELT$ & 0.05131 & 4.104 & 0.5492 & 0.2461 \\ 
    $\SCHACEcv$ & 0.0144 & \textbf{0.3215} & 0.1237 & \textbf{0.02418} \\ 
   \hline
\end{tabular}
\caption{Averaged mean squared errors for $n = 256$ and $a = 6$.} 
\label{tab:aMSEn256a6}
\end{table}

\begin{table}[ht]
\centering
\begin{tabular}{lcccc}
  \hline
Method & blocks & burt & cosine & heavisine \\ 
  \hline
$\pcs$ & 0.1616 &  4.33 & \textbf{1.017} & \textbf{0.2625} \\ 
  $\JIC$ & 1.875 & 28.93 & 2.387 & 1.723 \\ 
  $\PELT$ & \textbf{0.1444} &  12.5 & 1.399 & 0.8109 \\ 
    $\SCHACEcv$ & 0.2564 & \textbf{3.161} & 1.294 & 0.2634 \\ 
   \hline
\end{tabular}
\caption{Averaged mean squared errors for $n = 256$ and $a = 2$.} 
\label{tab:aMSEn256a2}
\end{table}

\begin{table}[ht]
\centering
\begin{tabular}{lcccc}
  \hline
Method & blocks & burt & cosine & heavisine \\ 
  \hline
$\pcs$ & \textbf{0.00449} & \textbf{0.3213} & \textbf{0.07179} & \textbf{0.02785} \\ 
  $\JIC$ & 1.318 & 14.91 & 0.1835 & 0.8053 \\ 
  $\PELT$ & 0.01354 & 2.331 & 0.2338 & 0.1573 \\ 
   \hline
\end{tabular}
\caption{Averaged mean squared errors for $n = 1024$ and $a = 4$.} 
\label{tab:aMSEn1024a4}
\end{table}

\begin{table}[ht]
\centering
\begin{tabular}{lcccc}
  \hline
Method & blocks & burt & cosine & heavisine \\ 
  \hline
$\pcs$ & \textbf{0.02761} & \textbf{1.118} & \textbf{0.3239} & \textbf{0.09617} \\ 
  $\JIC$ & 1.391 & 16.26 & 0.6343 & 0.8519 \\ 
  $\PELT$ & 0.03167 & 5.052 & 0.5094 & 0.3616 \\ 
   \hline
\end{tabular}
\caption{Averaged mean squared errors for $n = 1024$ and $a = 2$.} 
\label{tab:aMSEn1024a2}
\end{table}

\begin{table}[ht]
\centering
\begin{tabular}{lcccc}
  \hline
Method & blocks & burt & cosine & heavisine \\ 
  \hline
$\pcs$ & 0.1994 & \textbf{4.629} &  \textbf{1.09} & \textbf{0.2658} \\ 
  $\JIC$ & 1.622 & 21.94 & 2.426 &  1.03 \\ 
  $\PELT$ & \textbf{0.1701} & 12.14 & 1.181 & 0.8269 \\ 
   \hline
\end{tabular}
\caption{Averaged mean squared errors for $n = 1024$ and $a = 1$.} 
\label{tab:aMSEn1024a1}
\end{table}

\begin{table}[ht]
\centering
\begin{tabular}{llccccccccc}
 Setting & Method & $\hat{K} - K$ & $< -2$ & $-2$ & $-1$ & $0$ & $1$ & $2$ & $> 2$ & \% detected \\ 
 blocks & $\pcs$ & 0.8322 &  2.43 &  0.61 &  1.01 &  43.9 & 25.87 & 16.88 &   9.3 & 97.65 \\ 
  blocks & $\JIC$ & -5.63 &   100 &     0 &     0 &     0 &     0 &     0 &     0 & 48.76 \\ 
  blocks & $\PELT$ & 0.0925 &     0 &     0 &     0 &    \textbf{92} &  6.91 &  0.95 &  0.14 &   100 \\ 
  blocks & $\COPS$ & -6.71 &  99.8 &  0.07 &  0.13 &     0 &     0 &     0 &     0 & 37.45 \\ 
  blocks & $\SOPS$ & -9.299 &  98.5 &  0.49 &  0.61 &  0.17 &  0.11 &  0.08 &  0.04 & 12.38 \\ 
    blocks & $\SCHACEcv$ &  5.91 &     0 &     0 &     0 &  0.03 &  0.05 &  0.05 &  0.87 & 98.55 \\ 
   \hline
burt & $\pcs$ & 1.478 &     0 &     0 &  1.36 & 15.77 & 41.36 &  34.8 &  6.71 & 98.63 \\ 
  burt & $\JIC$ & 6.619 &     0 &     0 &  1.61 &  8.22 &  5.55 &  5.61 & 79.01 & 98.39 \\ 
  burt & $\COPS$ & 0.3373 &     0 &     0 &     0 & \textbf{81.33} &  9.96 &  4.98 &  3.73 &   100 \\ 
  burt & $\SOPS$ & -0.0173 &     0 &     0 & 83.01 &     0 &     0 &     0 & 16.99 & 16.99 \\ 
    burt & $\SCHACEcv$ &  0.73 &     0 &     0 &     0 &    70 &    15 &     8 &     7 &   100 \\ 
      \hline 
  cosine & $\pcs$ & 1.022 &   3.6 &  7.68 & 11.64 & 14.49 & 18.68 & 22.44 & 21.47 & 83.09 \\ 
cosine & $\JIC$ & 0.5833 &  0.15 &  2.85 & 15.18 & \textbf{40.18} & 20.98 & 10.77 &  9.89 &  92.7 \\ 
  cosine & $\COPS$ & -1.042 &   0.3 & 18.12 & 67.91 & 13.23 &  0.18 &  0.15 &  0.11 & 68.22 \\ 
  cosine & $\SOPS$ & -3.885 & 98.04 &     0 &     0 &     0 &  0.91 &  0.43 &  0.62 & 1.377 \\ 
      cosine & $\SCHACEcv$ &  8.22 &     1 &    10 &    12 &     2 &     1 &     4 &    70 & 84.75 \\ 
     \hline
  heavisine & $\pcs$ & -0.4368 &     0 & 49.05 & 17.69 & 14.86 &  5.94 &  2.62 &  9.84 & 32.83 \\ 
  heavisine & $\JIC$ & -0.0233 &     0 & 22.88 & 35.74 & 12.73 &  8.94 &  6.57 & 13.14 & 48.95 \\ 
heavisine & $\COPS$ & -0.359 &     0 &     0 & 56.02 & \textbf{30.95} &  8.42 &  2.83 &  1.78 & 65.21 \\ 
  heavisine & $\SOPS$ & -1.197 &     0 & 85.69 &     0 &     0 &     0 &     0 & 14.31 & 10.21 \\ 
    heavisine & $\SCHACEcv$ &  2.39 &     0 &     1 &    11 &    16 &    27 &    12 &    33 &    79 \\ 
  \end{tabular}
\caption{Summary results about the detection of change-points for $n = 256$ and $a = 4$.} 
\label{tab:CPn256a4}
\end{table}

\subsection{Application}\label{appendix:furtherApplication}
\begin{figure}[!htb]
\includegraphics[width = 0.9\textwidth]{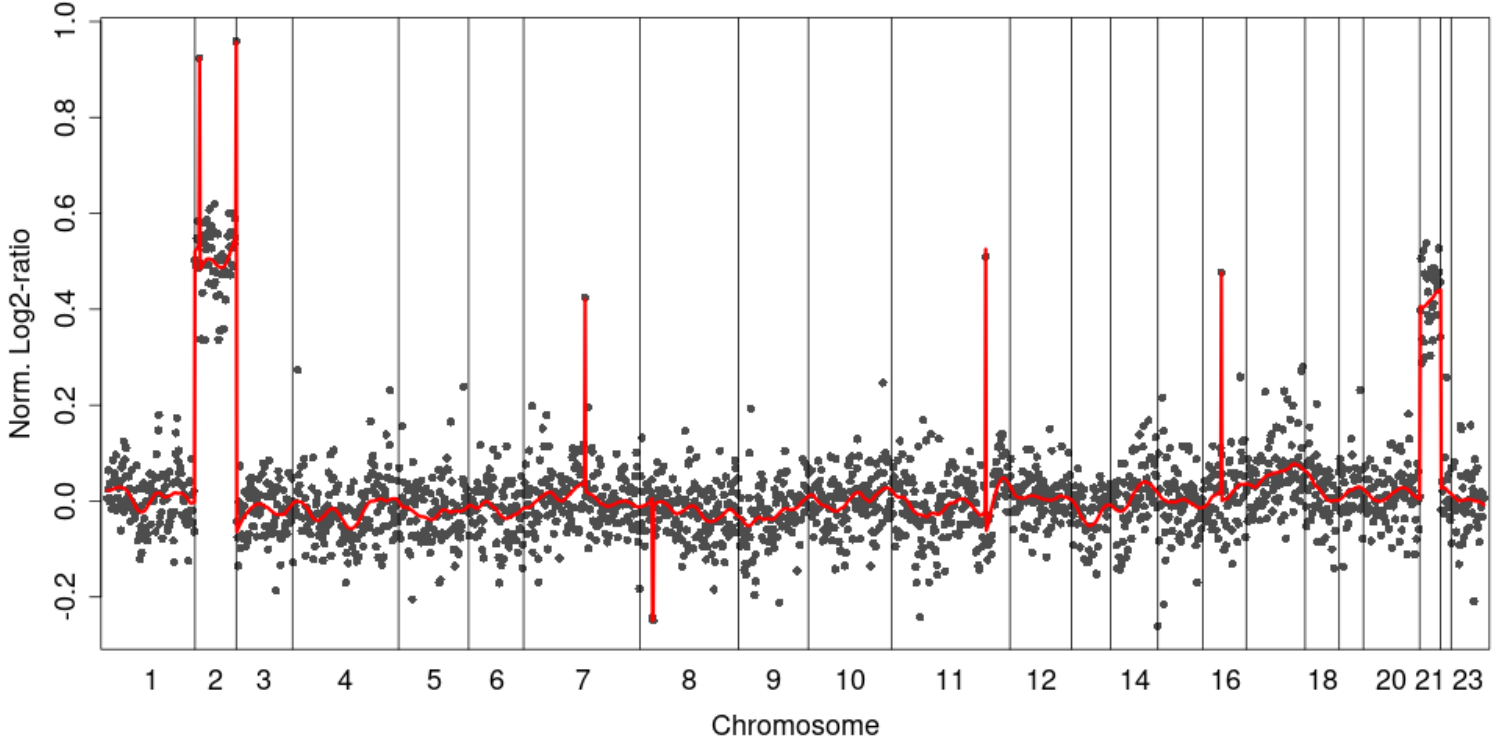} 
\caption{Cell line GM03576 analysed by $\PCpluS$ with slightly modified tuning parameters. It does not detect the additional false positives next to existing events.}
\label{fig:GM03576tuning}
\end{figure}

Figure \ref{fig:GM03576tuning} shows that $\PCpluS$ does not detect false positives next to the existing events if we decrease the $\lambda$ penalty by a factor of $0.65$.

\end{document}